\documentclass[11pt,a4paper]{article}
\pdfoutput=1
\usepackage{jheppub}
\usepackage{amsmath}
\usepackage{verbatim}
\usepackage{graphicx}
\usepackage{mathrsfs}
\usepackage{appendix}
\usepackage{caption}
\usepackage{float}
\usepackage{subfig}
\usepackage{mathtools}
\usepackage{jheppub}
\usepackage{amsfonts}
\usepackage{enumerate}
\definecolor{mycolor}{RGB}{24, 116, 148}

\usepackage{amssymb}
\usepackage{inputenc, array}
\usepackage{textcomp}
\usepackage[dvipsnames]{xcolor}

\newcommand{\be}[1]{ \begin{equation}\label{#1} }
\newcommand{\ee}{\end{equation}}
\newcommand{\bea}[1]{\begin{eqnarray}\label{#1} }
\newcommand{\eea}{\end{eqnarray}}
\newcommand{\bes}{\begin{subequations}}
\newcommand{\ees}{\end{subequations}}

\definecolor{mycolor}{RGB}{24, 116, 148}
\hypersetup{colorlinks=true,citecolor=mycolor,linkcolor=mycolor, linktocpage,urlcolor=mycolor}

\renewcommand{\S}{\mathcal{S}}

\title{Quantum corrections to the Near-Extremal Thermodynamics of (warped) BTZ Black Holes}

\author{Emilie Despontin,} \author{St\'ephane Detournay,} \author{Robinson Mancilla} \author{and Chiara Toldo \\}

\affiliation{Physique Th\'eorique et Math\'ematique and International Solvay Institutes,\\
Universit\'e Libre de Bruxelles (ULB), C.P. 231, 1050 Brussels, Belgium.}
\emailAdd{emilie.despontin@ulb.be, stephane.detournay@ulb.be, robinson.mancilla.perez@ulb.be, chiara.toldo@ulb.be}

\abstract{We study one-loop effects in the near-extremal thermodynamics of BTZ and warped BTZ black holes, with particular emphasis on the fate of eigenmodes that become zero modes in the extremal throat. Our analysis is formulated in three-dimensional Topologically Massive Gravity, a higher derivative theory characterized by the presence of a gravitational Chern--Simons term, and it makes use of the Newman--Penrose formulation.  For BTZ, we compare the near-horizon computation with the full-geometry eigenvalue problem and identify how the Schwarzian and rotational sectors are lifted at small temperature. We then extend the same strategy to warped BTZ. We find that rotational modes are essential for a consistent near-extremal treatment of both BTZ and warped BTZ black holes, and cannot be discarded without first specifying the boundary conditions.}


\preprint{}

\begin{document}
\maketitle

\section{Introduction }

The gravitational path integral is one of the few frameworks in which black hole thermodynamics can be treated as a first-principles problem in quantum gravity. Given suitable boundary conditions, one formally defines a partition function by summing over bulk geometries and matter configurations compatible with those data. In the semiclassical expansion, this sum is organized by classical saddles, their boundary terms, and the determinants of quadratic fluctuations around each saddle. This perspective determines which configurations are counted, and which apparent classical degeneracies survive quantization. These questions become especially sharp near extremality, where the leading saddle approximation predicts a macroscopic zero-temperature entropy but the quantum fluctuations 
can qualitatively change the low-temperature physics \cite{Iliesiu:2020qvm,Heydeman:2020hhw,Iliesiu:2022onk,Mertens:2022irh,Turiaci:2023wrh}.\\

Three-dimensional gravity provides a particularly stringent arena for this problem. Pure Einstein gravity with negative cosmological constant has no local propagating gravitons, yet it has black holes \cite{Banados:1992wn,Banados:1992gq}, boundary gravitons, a Brown--Henneaux asymptotic symmetry algebra \cite{Brown:1986nw}, and a non-trivial Euclidean path integral over geometries with torus boundary \cite{Witten:2007kt}. The resulting partition function is therefore simple enough to be attacked directly and subtle enough to test the meaning of quantum gravity. In the classic analysis of pure three-dimensional quantum gravity, the known saddle contributions can be evaluated, including quantum corrections, but the resulting partition function is not by itself physically sensible, suggesting that pure three-dimensional gravity either requires additional contributions or does not exist as a standalone non-perturbative theory in the naive form \cite{Maloney:2007ud,Cotler:2020hgz,Belin:2023efa}. At the perturbative level, the one-loop determinants on locally AdS$_3$ geometries can be computed explicitly by heat-kernel methods. For the graviton, the result reproduces the Virasoro boundary excitations expected from the Brown--Henneaux analysis and gives the one-loop correction to thermal AdS$_3$, BTZ, and related quotient geometries \cite{Giombi:2008vd}.\\

Topologically Massive Gravity  enriches this laboratory while preserving much of its analytic control. The addition of the gravitational Chern--Simons term introduces a parity-violating massive spin-two mode and allows for a richer variety of geometries and boundary conditions.
At the chiral point, the distinction between chiral gravity and logarithmic gravity is tied to linearization instabilities, logarithmic modes, and the possible dual interpretation in terms of extremal or logarithmic CFTs \cite{Maloney:2009ck}. The AdS$_3$ one-loop partition function of TMG  reflects these features: it is not holomorphically factorized in general, and at the chiral point it has the structure expected of a logarithmic conformal field theory \cite{Gaberdiel:2010xv}. Moreover, away from the locally AdS$_3$ sector, TMG admits warped AdS$_3$ vacua and warped black holes, which may be viewed as discrete quotients of warped AdS$_3$ in close analogy with the relation between BTZ and AdS$_3$ \cite{Anninos:2008fx,Aggarwal:2023peg}. In addition, warped AdS$_3$ geometries play a role in the physics of Kerr black holes \cite{Guica:2008mu} since they arise from the Near Horizon Extremal Kerr (NHEK) metric \cite{Bardeen:1999px} at fixed polar angle. 
These facts make TMG and warped black holes a natural setting in which to ask how the gravitational path integral behaves when contributions arise simultaneously from the near-horizon throat, the asymptotic region, and the massive graviton sector, for geometries closer in spirit to four-dimensional extremal black holes.\\

In this paper we investigate the presence of zero modes responsible for $3/2 \log T$ corrections \cite{Iliesiu:2022onk} in the near-horizon geometry of (warped) BTZ in TMG. In the strict extremal
limit, the near-horizon geometry contains an AdS$_2$ factor, and admits normalizable
reparametrization modes. The path integral over these modes is infrared
divergent at zero temperature. Keeping the leading finite-temperature
deformation of the throat regulates this divergence by lifting the zero
modes, producing a characteristic universal
$T^{3/2}$ prefactor in the canonical partition function \cite{Iliesiu:2022onk}, which was found also for the BTZ geometry in \cite{Kolanowski:2024zrq,Kapec:2024zdj}. For BTZ the $3/2\log T$ scaling can also be obtained by modular invariance of the dual 2d  CFT partition function in a near-extremal regime \cite{Ghosh:2019rcj}.\\

One of the main purposes of this paper is to connect the throat analysis to computations performed in the full geometry. While the near-horizon region is often expected to capture the relevant low-temperature physics, especially in holographic settings, the associated degrees of freedom should ultimately admit a description in the boundary theory. Relating near-horizon modes to their full-geometry counterparts has therefore been the subject of a growing body of recent work, see for example  \cite{Kolanowski:2024zrq,Acito:2025hka,Bac:2026eqj}. In the particular case of BTZ, the full-geometry analyses   have identified off-shell eigenmodes of the Euclidean quadratic fluctuation operator whose eigenvalues vanish linearly with temperature and whose wavefunctions localize near the horizon as extremality is approached\footnote{The $T^{3/2}$ scaling can also be obtained using the Denef-Hartnoll-Sachdev formula \cite{Denef:2009kn,Denef:2009yy}  of one-loop determinants in terms of quasinormal modes \cite{Kapec:2024zdj,Arnaudo:2024bbd,Arnaudo:2025btb}.}.  \\

By extending the near-extremal one-loop analysis to BTZ and warped BTZ black holes in TMG, we aim to clarify how universal Schwarzian physics is embedded in the full three-dimensional gravitational path integral, and how it is modified when the theory contains parity-violating massive gravitons and warped asymptotics.
 One of the tools at our disposal, is the Newman--Penrose (NP) formalism \cite{Newman:1961qr}, which allows us to rewrite the equations of motion using a set of null vectors instead of the usual coordinate basis. This formalism, locally valid for any sufficiently smooth four-dimensional Lorentzian spacetime, independently of the specific gravitational field equations, allowed us to construct the modes in the full geometry from the near-horizon ones, potentially paving the way to a higher dimensional generalization of our procedure. \\ 
 
 The findings of our paper can be summarized as follows: 
\begin{itemize}
\item Extremal (warped) BTZ black holes in TMG admit an infinite family of normalizable zero modes $h_{\mu \nu}$ in their near-horizon geometry. These correspond to diffeomorphisms left unfixed by the gauge choice and they consist of tensor (Schwarzian) modes and U(1) (rotational) modes. Using an appropriately regulated near-extremal geometry, we find their eigenvalue corrections $\delta \lambda_{\rm Schw.}$ and $\delta \lambda_{\rm Rot.}$. These lead to well-defined $\log T$-corrections in the path integral.

\item These modes have a specific decomposition in terms of Newman--Penrose  triads in the near-horizon geometry, which we detail in Secs. \ref{sec:NHBTZ} and \ref{sec:ThroatAnalysis}.
 
\item The full non-extremal black hole geometry admits a finite set of normalizable ``off-shell" modes which are eigenmodes of the TMG Lichnerowicz operator
\begin{equation}\label{lichn}
\mathcal{L}^{\rm TMG} h_{\mu \nu} = \lambda (T) h_{\mu \nu}.
\end{equation}
To find these modes, we used the same ansatz and NP decomposition, using the triads \textit{in the full geometry}. These modes become infinite in number in the extremal limit, and a subset of them coincide with the near-horizon zero modes.

\item For the Schwarzian modes, the limit for $T\rightarrow 0$ of the full geometry eigenmodes and of $\lambda(T)$ coincides exactly with the value found in the near-horizon geometry $\delta \lambda_{\rm Schw.}$.  

\item For the rotational modes, the situation is more subtle. The full geometry modes contain a component which turns out to be non-normalizable in the near-horizon limit. This makes the limit $T \rightarrow 0$ of the exact $\lambda(T)$ naively disagree with the one found in the near-horizon geometry $\delta \lambda_{\rm Rot.}$. This discrepancy arises from a subtlety regarding perturbation theory when applied in the near-horizon region, and in particular to the fact that for the transverse-traceless gauge to be satisfied in the regulated geometry such non-normalizable component is indeed necessary. The discrepancy is then resolved by a careful treatment of perturbation theory in the near-horizon region. 

\item The Schwarzian modes in the full BTZ geometry have fall-offs that are incompatible with standard Brown–Henneaux boundary conditions \cite{Brown:1986nw} at finite temperature. However, in the very low-temperature regime, their asymptotic behaviour becomes compatible with the near-horizon analysis. This suggests an order-of-limits issue between imposing boundary conditions at infinity and taking the near-extremal limit. For warped BTZ black holes, an analogous subtlety arises: the Schwarzian and rotational modes are not part of the finite-temperature phase space for arbitrary mode number, but become compatible with the relevant boundary conditions in the near-extremal limit.

\end{itemize}
 
Our findings need to be compared with recent work on the near-extremal limit of the field theory dual, which is a warped conformal field theory (WCFT) \cite{Aggarwal:2022xfd,Aggarwal:2023peg}, invariant under
one copy of the Virasoro algebra, and one $\hat{u}(1)$ Kac–Moody algebra. As we will see, some subtleties arise concerning the choice of ensemble and the sign of the eigenvalue correction.\\

The paper is organized as follows. In Section \ref{TMG_sec}, we detail the framework of Topological Massive Gravity and we further spell out our framework for the computation of the one-loop determinant. We moreover provide a simplified form for the Lichnerowicz operator and we point out subtleties arising with perturbation theory. We apply this framework to the BTZ geometry in TMG in Section \ref{sec:BTZinTMG}. There, we spell out the zero modes in the near-horizon geometry, the quantum corrected eigenvalues, and the full-geometry extension. All this is generalized to the warped BTZ black hole in TMG in Section \ref{sec:WBTZ}. In the conclusions we elaborate on dual WCFT picture and we give perspectives for future work.\\

\section{TMG, Gauge Fixing and Perturbation Theory\label{TMG_sec}}

In this section, we introduce Topological massive gravity (TMG) and its linearized theory. After this, we simplify the Lichnerowicz operator acting on transverse-traceless sector and discuss the associated gauge fixing. Finally, we turn to perturbation theory, where we introduce our prescription for handling a certain classes of non-normalizable modes within the black hole throat near-extremality. 

\subsection{Topological Massive Gravity\label{TMG_intro}}
Topologically massive gravity is obtained by supplementing the Einstein--Hilbert action with a gravitational Chern--Simons term \cite{Deser:1982TMG,Deser:1982TMG2,Deser:1991qk,Deser:2002iw}. While pure three-dimensional Einstein gravity has no local propagating graviton degrees of freedom, TMG supports a massive bulk graviton, thereby providing a setting in which genuinely dynamical gravitational effects can be studied without losing the analytic control characteristic of three dimensions \cite{Li:2008dq}. Concretely, the action of three-dimensional TMG with negative cosmological constant $\Lambda = -1/\ell^{2}$ is given by the Einstein--Hilbert action 
\begin{equation}
    \S_{\rm EH} = \frac{1}{2\kappa^2}\int d^3x\, \sqrt{-g} \left(R+\frac{2}{\ell^2}\right),
\end{equation}
together with the gravitational Chern--Simons term
\begin{equation}
    \S_{\rm CS} = \frac{1}{4\mu\kappa^2}\int d^3x \, \sqrt{-g}\,\varepsilon^{\mu\nu\rho} \left(\Gamma^\alpha{}_{\mu\beta}\partial_{\nu}\Gamma^\beta{}_{\rho\alpha}+ \frac23 \Gamma^\kappa{}_{\mu\beta}\Gamma^\beta{}_{\nu\alpha}\Gamma^\alpha{}_{\rho\kappa}\right),
\end{equation}
such that
\begin{equation}
\label{eq:TMGaction}
    \S_{\,\rm TMG} = \S_{\, \rm EH} + \frac 1 \mu \S_{\, \rm CS}.
\end{equation}
The parameter $\mu$ is a real positive coupling constant that sets the mass of the propagating graviton \cite{Li:2008dq}, while $\ell$ denotes the AdS radius. The equations of motion are
\begin{align}
    R_{\mu\nu}-\frac{1}{2}g_{\mu\nu}R-\frac{1}{\ell^{\, 2}}g_{\mu\nu}+\frac{1}{\mu}C_{\mu\nu}=0,
\end{align}
where $C_{\mu\nu}$ is the Cotton tensor, defined by the curl of the Schouten tensor $S_{\mu\nu}$
\begin{equation}
    C_{\mu\nu}=D_{\mu}{}^{\beta}S_{\beta\nu},\qquad
    S_{\beta\nu}=R_{\beta\nu}-\frac{1}{4}g_{\beta\nu}R,
\end{equation}
where $\varepsilon_{\mu\nu\alpha}$ is the Levi-Civita tensor\footnote{The fully anti-symmetric tensor is defined as $\varepsilon_{\mu\nu\alpha}=\sqrt{-g}\epsilon_{\mu\nu\alpha}$ where $\epsilon_{\mu\nu\alpha}$ is the Levi-Civita symbol with the convention $\epsilon_{012}=-1$. Later, we consider the Wick rotation $t\to-i\tau$ where $t$ is Lorentzian time, and $\tau$, Euclidean time. Then, our convention is $\epsilon_{012}=i$. In words, the Levi-Civita symbol contains the factor of $i$.} and $D_{\mu}{}^{\beta}$ stands for the curl covariant derivative defined as  \cite{Tyutin:1997yn}
\begin{equation}
\label{eq:curlD}
    D_{\mu}{}^{\beta}\equiv \varepsilon_{\mu}{}^{\alpha\beta}\,\nabla_{\alpha}.
\end{equation}
The Cotton tensor satisfies the following properties 
\begin{equation}
    C=0,\qquad \nabla^{\mu}C_{\mu\nu}=0,\qquad C_{\mu\nu}=C_{\nu\mu}.
\end{equation}
Despite the presence of the Cotton tensor, all TMG solutions have negative Ricci scalar $R=-6/\ell^2$.  Inspired by this, one introduces the traceless Ricci tensor, defined as
\begin{equation}
\label{eq:Edef}
    E_{\mu\nu}=R_{\mu\nu}-\frac{1}{3}g_{\mu\nu}R,
\end{equation}
which allows to recast the TMG equations of motion as follows
\begin{equation}
\label{eq:TMGeom2}
    \left(\delta_{\mu}^{\beta}+\frac{1}{\mu}D_{\mu}{}^{\beta}\right)E_{\beta\nu}=\frac{g_{\mu\nu}}{6}\left(R+\frac{6}{\ell^2}\right)-\frac{g_{\nu\beta}}{12}D_{\mu}{}^{\beta}R.
\end{equation}
Indeed, the right-hand side of this expression vanishes on shell, since
$R=-6/\ell^2$. This is precisely the form used in \cite{Garcia:2003bw,Lancaster:1986epo,Chow:2009km} to study the Petrov-Segre algebraic classification of spacetime solutions in TMG. Later, we will use this form of the equations of motion to analyze linear fluctuations, since it makes it easier to decouple the trace mode from the transverse-traceless mode by exploiting the algebraic properties of the background solutions.

\paragraph{Locally AdS$_3$ and warped backgrounds.}\label{Sec:LocallyAdSWBTZ} In TMG, spacetime solutions can be classified into those with vanishing traceless Ricci tensor, $E_{\mu\nu}=0$, and those with a non-trivial traceless Ricci tensor, $E_{\mu\nu}\neq 0$. In particular, the BTZ black hole is a solution of TMG. However, its mass, angular momentum, and entropy differ from their values in pure AdS$_3$ gravity, since these quantities receive contributions from the gravitational Chern--Simons term. Moreover, in the locally AdS$_3$ classical phase space of TMG, imposing Brown--Henneaux boundary conditions leads to a putative dual CFT that is parity-violating, since the left- and right-moving central charges are different, $c_{L}\neq c_{R}$ \cite{Kraus:2005zm,Solodukhin:2005ns,Solodukhin:2006chern,Hotta:2008yq,Skenderis:2009nt}. Furthermore, it is also possible to impose the CSS boundary conditions \cite{Compere:2013bya,Aggarwal:2020igb} on this phase space, leading to a putative warped CFT \cite{Detournay:2012pc,Ciambelli:2020shy}. We study this case in Section \ref{sec:BTZinTMG}, where further details are provided.\\

On the other hand, warped backgrounds have a non-vanishing traceless Ricci tensor and are intrinsic solutions of TMG. An important subclass consists of the WAdS$_3$ vacuum, warped black holes, and self-dual WAdS$_3$ spacetimes \cite{Li:2008dq,Anninos:2008fx}, which are type-$D_s$ solutions in the Petrov-Segre algebraic classification \cite{Chow:2009km}. These geometries admit a unit-normalized spacelike Killing vector $p^\mu$ such that the traceless Ricci tensor takes the type-$D_s$ form
\begin{equation}
\label{eq:backgroundE}
    E_{\mu\nu} = \frac{\nu^2-1}{\ell^2}\left(g_{\mu\nu} - 3 p_\mu p_\nu \right),
\end{equation}
where the parameter $\nu$ is fixed by the equations of motion in terms of $\mu$ as
\begin{equation}
    \nu = \frac{\mu\ell}{3}.
\end{equation}
In order to avoid closed timelike curves in warped black holes, we assume that $\nu^2\geq1$ \cite{Anninos:2008fx}. The locally AdS$_3$ solutions are recovered in the limit $\nu\to1$, corresponding to the parameter choice $\mu=3/\ell$. However, we emphasize that locally AdS$_3$ solutions exist for arbitrary $\mu$. In Section \ref{sec:WBTZ}, we will discuss the warped BTZ solution, where the Killing vector $p^\mu$ will be given explicitly.\\

\subsection{Lichnerowicz Operator}
\label{sec:WBTZLICh}
Within this framework, we are interested in computing the gravitational path integral 
\begin{equation}\label{PI}
    \mathcal Z = \int \mathcal D g \, e^{-\mathcal S_{\rm TMG}},
\end{equation}
where we sum over geometries that satisfy certain boundary conditions fixed by thermodynamic ensembles, and prescribed metric fall-offs. We will compute the gravitational path integral in the saddle point approximation, namely using the expansion of the metric around a solution of the equations of motion with metric $\bar{g}$, given  by
\begin{equation} \label{expansion}
    g_{\mu\nu} = \bar g_{\mu\nu} + h_{\mu\nu}.
\end{equation}
In the saddle point approximation, Eq. \eqref{PI} can be rewritten as
\begin{equation}
    \mathcal Z \sim e^{-\mathcal S_{\rm TMG}[\bar{g}]}\int \mathcal{D}h \exp \left(-\int d^3x \, \sqrt{\bar{g}}\, h^{\mu\nu}\mathcal{L}_{\mu\nu,\alpha\beta} h^{\alpha\beta}\right).
\end{equation}
The differential operator $\mathcal{L}$, which we keep generic for the moment, is called the \textit{Lichnerowicz operator}. Computing the one-loop correction to the path integral therefore amounts to evaluating the functional determinant of this operator 
\begin{equation}
    \log \mathcal{Z}_{\rm one-loop} = -\frac12 \log \det \mathcal{L}.
\end{equation}
The spectrum of the Lichnerowicz operator therefore determines the one-loop correction. \\

We will focus first in the gravitational path integral in the near-horizon geometry, as done in \cite{Sen:2012cj,Sen:2012dw}. In particular, as we discuss below, near the extremal horizon the presence of \textit{zero modes}, namely eigenmodes of $\mathcal L$ with vanishing eigenvalue, leads to an infrared divergence that must be regularized. The backgrounds considered here possess an infinite tower of such zero modes. These originate from metric fluctuations left unfixed by the gauge fixing, in particular from boundary time diffeomorphisms (acting on the AdS$_2$ part of the geometry) and modes deforming the $U(1)$ fiber.  To render the path integral finite, we consider a slight departure from extremality obtained by exciting the black hole to a small temperature \cite{Iliesiu:2022onk}. The deformation lifts the zero modes, replacing their vanishing eigenvalues with small nonzero ones, and the resulting quantum contribution dominates the path integral in the near-extremal limit.\\

First of all, we will compute the Lichnerowicz operator in TMG and massage its expression to put it in a more convenient form. The quadratic action, obtained from the TMG action \eqref{eq:TMGaction} upon using \eqref{expansion}, then takes the form
\begin{equation}
    \mathcal S_{\rm TMG}^{(2)}=\frac{1}{32\pi G_3}\int d^3x \sqrt{\bar{g}}\, h^{\mu\nu}  \left(\mathcal{L}^{\rm TMG}h\right)_{\mu\nu},
\end{equation}
where the TMG Lichnerowicz operator $\mathcal{L}^{\rm TMG}$ is given by
\begin{align}
    (\mathcal{L}^{\rm TMG}h)_{\mu\nu}=&\left(\delta_{\mu}^{\alpha}+\frac{1}{\mu}\bar{D}_{\mu}{}^{\alpha}\right)E^{(1)}_{\alpha\nu}-\frac{h}{2}\bar{E}_{\mu\nu} -\frac{\bar{g}_{\mu\nu}}{6}R^{(1)}
    +\frac{1}{12\mu}\bar{g}_{\nu\beta}\bar{D}_{\mu}{}^{\beta}R^{(1)}\nonumber\\
    &-\frac{\epsilon_{\mu}{}^{\alpha\beta}}{\mu}\left(\bar{E}_{\beta\lambda}\Gamma^{\lambda(1)}_{\alpha\nu}+h_{\alpha\lambda}\nabla^{\lambda}\bar{E}_{\beta\nu}+h_{\beta}{}^{\lambda}\nabla_{\alpha}\bar{E}_{\lambda\nu}\right),
\end{align}
where the linearized traceless Ricci tensor $E^{(1)}_{\mu\nu}$ is
\begin{equation}
    E^{(1)}_{\mu\nu}=R^{(1)}_{\mu\nu}+\frac{2}{\ell^2}h_{\mu\nu}-\frac{\bar{g}_{\mu\nu}}{3}R^{(1)}.
\end{equation}
The relevant linearized tensors are given by
\begin{align}
\Gamma^{\lambda(1)}_{\alpha\nu}&=\frac12\left(\bar\nabla_{\alpha}h^{\lambda}{}_{\nu}+\bar\nabla_{\nu}h^{\lambda}{}_{\alpha}-\bar\nabla^{\lambda}h_{\alpha\nu}\right),\\
    R^{(1)}_{\mu\nu}&=-\frac12\left(\bar\nabla_{\mu}\bar\nabla_{\nu}h-\bar\nabla_{\mu}\bar\nabla_{\rho}h^{\rho}{}_{\nu}-\bar\nabla_{\nu}\bar\nabla_{\rho}h^{\rho}{}_{\mu}+\bar\Box h_{\mu\nu}\right)
    -\bar R_{\alpha\mu\beta\nu}h^{\alpha\beta}+\bar R_{(\mu|\rho|}h^{\rho}{}_{\nu)},\\
    R^{(1)}&=\bar\nabla_{\rho}\bar\nabla_{\sigma}h^{\rho\sigma}-\bar\Box h-\bar R_{\rho\sigma}h^{\rho\sigma}.
\end{align}
The TMG action \eqref{eq:TMGaction} is not manifestly diffeomorphism invariant, since the gravitational Chern--Simons term is written in terms of the Christoffel connection. Nevertheless, the TMG equations of motion \eqref{eq:TMGeom2} are diffeomorphism invariant. Consequently, the quadratic action \eqref{eq:quadraticTMG1} is diffeomorphism invariant as well.\\

\paragraph{TT-gauge decomposition.} 
 We decompose the fluctuations into a traceless-transverse $\left(h^{\rm TT}\right)_{\mu\nu}$ mode, a traceless mode $h$, and a diffeomorphism as follows 
\begin{align}
    h_{\mu\nu}=\left(h^{\rm TT}\right)_{\mu\nu}+\frac{1}{3}\bar{g}_{\mu\nu}h+2\nabla_{(\mu}\xi_{\nu)},\label{Eq:TTDecomposition}
\end{align}
with $\left(h^{\rm TT}\right)_{\mu\nu}$ satisfying
\begin{equation}
      \left(h^{\rm TT}\right)^{\mu}{}_{\mu}=0,\qquad \nabla^{\mu}\left(h^{\rm TT}\right)_{\mu\nu}=0.
\end{equation}
We remark that the arbitrary diffeomorphism $\xi_{\mu}$ affects both the transverse-traceless mode $\left(h^{\rm TT}\right)_{\mu\nu}$ and the trace mode $h$. Given this decomposition, the quadratic action becomes
\begin{equation}
\label{eq:quadraticTMG1}
   \S^{(2)}_{\rm TMG}=-\frac{1}{2\kappa^2}\int d^3x \,\sqrt{g}\, \Big[ \left(h^{\rm TT}\right)^{\mu\nu}\left(\mathcal{L}h^{\rm TT}\right)_{\mu\nu}+\frac{1}{9}h\left(\Box-\frac{3}{\ell^2}\right)h+\frac{4h}{3}\; (h^{\rm TT})^{\mu\nu}\bar{E}_{\mu\nu}\Big].
\end{equation}
Notice that, due to gauge invariance, the field $\xi_{\mu}$ does not appear. The Lichnerowicz operator acting on TT-mode in the type-$D_s$ class of solutions defined by Eq.  \eqref{eq:backgroundE} is 
\begin{align}
\label{eq:lichTMGTT}
    \left(\mathcal{L}h^{\rm TT}\right)_{\mu\nu}&=\left(\delta_{\mu}^{\alpha}+\frac{1}{\mu}\bar{D}_{\mu}{}^{\alpha}\right)E^{(1)}_{\alpha\nu}[h^{\rm TT}]\nonumber\\
    &-\frac{\epsilon_{\mu}{}^{\alpha\beta}}{\mu}\left(\bar{E}_{\beta\lambda}\Gamma^{\lambda(1)}_{\alpha\nu}[h^{\rm TT}]+\left(h^{\rm TT}\right)_{\alpha\lambda}\nabla^{\lambda}\bar{E}_{\beta\nu}+\left(h^{\rm TT}\right)_{\beta\lambda}\nabla_{\alpha}\bar{E}^{\lambda}{}_{\nu}\right),\\
    E^{(1)}_{\mu\nu}[h^{\rm TT}]&=\frac{1}{2}\left(-\Box-\frac{2}{\ell^2}\right)\left(h^{\rm TT}\right)_{\mu\nu}+3\bar{E}_{\alpha(\mu}\left(h^{\rm TT}\right)_{\nu)}{}^{\alpha}.
\end{align}
We observe that there is an explicit interaction between the transverse-traceless mode $h^{\rm TT}_{\mu\nu}$ and the trace mode $h$ through the background traceless Ricci tensor $\bar{E}_{\mu\nu}$. For locally AdS$_3$ solutions, where $\bar{E}_{\mu\nu}=0$, we recover the Lichnerowicz operator used in \cite{Gaberdiel:2010xv} to study the one-loop partition function of this class of solutions in TMG.\\

We can diagonalize the quadratic action \eqref{eq:quadraticTMG1} in field space by introducing the field redefinitions
\begin{align}
    &F^{\pm}_{\mu\nu}\equiv \frac{1}{\sqrt{2}}\left((h^{\rm TT})_{\mu\nu}\pm \frac{1}{3}\bar{g}_{\mu\nu}h\right),\\
    &\mathcal{E}^{\alpha\beta}_{\mu\nu}\equiv \frac{4}{3}\bar{g}^{\alpha\beta}\bar{E}_{\mu\nu}.
\end{align}
In terms of these new fields, the quadratic action takes the form
\begin{equation}
   \S^{(2)}_{\rm TMG}=-\frac{1}{2\kappa^2}\int d^3x \sqrt{g}\; \Big[ F^{+}(\mathcal{L}+\mathcal{E})F^{+}+F^{-}(\mathcal{L}-\mathcal{E})F^{-}\Big],
\end{equation}
where the index contractions are kept implicit. Thus, the one-loop partition function is
\begin{equation}
    \mathcal Z_{\rm one-loop}^{\rm TMG}=\frac{\mathcal Z_{\rm ghost}}{\sqrt{\rm det (\mathcal{L}+\mathcal{E})_{+}  det (\mathcal{L}-\mathcal{E})_{-}}}.
\end{equation}
If one wants to study the one-loop partition function around a warped black hole in TMG, this diagonalization in field space is certainly desirable. For a type-$D_s$ background defined in \eqref{eq:backgroundE}, the interaction term reduces to
\begin{equation}
\frac{4h}{3}\left(h^{\rm TT}\right)_{\mu\nu}\bar{E}^{\mu\nu}=-\frac{4(\nu^2-1)}{\ell^2}\left(h^{\rm TT}\right)_{\mu\nu}p^{\mu}p^{\nu}\;h .
\end{equation}
We will see later that the Schwarzian and rotational modes, which are the modes of interest for this work, satisfy
\begin{equation}
   \left(h^{\rm TT}\right)_{\mu\nu}p^{\mu}p^{\nu}=0.
\end{equation}
Thus, the coupling between the transverse-traceless mode $\left(h^{\rm TT}\right)_{\mu\nu}$ and the trace mode $h$ will not play any role in our discussion. Nevertheless, for completeness, in the following section we discuss the ghost contribution $\mathcal Z_{\rm ghost}$ for arbitrary metric fluctuations.\\

\subsection{Gauge Fixing and Ghosts}
Since TMG contains third-order derivatives of the metric due to the gravitational Chern--Simons term, the corresponding gauge-fixing condition should also involve third-order derivatives of the metric. To avoid this technical complication, we follow the procedure of \cite{Gaberdiel:2010xv}, which is based on an explicit separation of the gauge modes, to derive the ghost contribution $\mathcal Z_{\rm ghost}$ to the partition function in a more efficient way. We briefly review its main features here. The diffeomorphism parameter $\xi_\mu$ introduced in Eq. \eqref{Eq:TTDecomposition} can be further decomposed into a transverse part $\xi_\mu^{\rm T}$ and a scalar part $\sigma$
\begin{equation}
    \xi_\mu = \xi_\mu^{\rm T} + \nabla_\mu \sigma,
\end{equation}
where the transverse condition is
\begin{equation}
    \nabla^\mu \xi_\mu^{\rm T} =0.
\end{equation}
After the field redefinition $h\to h - 2 \Box \sigma$, leaving the resulting trace of $h_{\mu\nu}$ independent of $\xi^{\rm T}_\mu$, we derive the following decomposition for the metric fluctuation
\begin{equation}
\label{eq:hdecomposition2}
    h_{\mu\nu} = h_{\mu\nu}^{\rm TT} + \frac13 \bar g_{\mu\nu} h+ 2\nabla_{(\mu}\xi_{\nu)}^{\rm T} + 2 \nabla_\mu\nabla_\nu \sigma - \frac23 g_{\mu\nu} \Box \sigma.
\end{equation}
To determine the associated Jacobian, we use the Gaussian normalization of the path integral measures defined as
\begin{align}
    \int \mathcal{D}h_{\mu\nu}\,
        e^{-\left\langle  h|h\right\rangle}
        &= 1,
    &
    \left\langle  h|h'\right\rangle
        &= \int d^3x\,\sqrt{g}\,
        h^{\mu\nu}h'_{\mu\nu},
    \\
    \int \mathcal{D}\xi_{\mu}\,
        e^{-\left\langle  \xi|\xi\right\rangle}
        &= 1,
    &
    \left\langle  \xi|\xi'\right\rangle
        &= \int d^3x\,\sqrt{g}\,
        \xi^{\mu}\xi'_{\mu},
    \\
    \int \mathcal{D}\sigma\,
        e^{-\left\langle  \sigma|\sigma\right\rangle}
        &= 1,
    &
    \left\langle  \sigma|\sigma'\right\rangle
        &= \int d^3x\,\sqrt{g}\,
        \sigma\sigma'.
\end{align}
We note that each term in the fluctuation decomposition \eqref{eq:hdecomposition2} is orthogonal to each other with respect to the ultra-local measure introduced above. This allows us to compute the Jacobians associated with the change of variables in the path integral measure in an efficient manner. Let us first compute the Jacobian $J_1$ associated with the decomposition of the diffeomorphism parameter $\xi_{\mu}$. One has
\begin{align}
    1&=\int \mathcal{D}\xi_{\mu}^{\rm T} \mathcal{D}\sigma J_1 \;e^{-\left\langle  \xi|\xi\right\rangle}=\int \mathcal{D}\xi_{\mu}^{\rm T} \mathcal{D}\sigma \, J_1 \exp\left[\int d^3x \sqrt{g} \left(\sigma\Box\sigma-\xi^{\rm T}_{\mu}\xi^\mu_{\rm T}\right)\right] =J_1/  \sqrt{\det(-\Box)_0 },
\end{align}
resulting in 
\begin{equation}
   J_1=\sqrt{\det(-\Box)_0 }.
\end{equation}
In a similar manner, after a few integrations by parts and using the background traceless Ricci tensor $\bar{E}^{\mu\nu}$ introduced in Eq. \eqref{eq:backgroundE}, we obtain the Jacobian $J_2$ associated with the decomposition of the fluctuation $h_{\mu\nu}$ in the path-integral measure
\begin{align}
    1&=\int \mathcal{D}\left(h^{\rm TT}\right)_{\mu\nu} \mathcal{D}\sigma \mathcal{D}h \mathcal{D}\xi^{\rm T}_{\mu} J_2\; e^{-\left\langle  h|h\right\rangle},
\end{align}
resulting in
\begin{equation}
    J_2=\sqrt{\det(-\mathcal{A})_0\det(-\mathcal{B})_{1}},
\end{equation}
where the operators $\mathcal{A}$ and $\mathcal{B}$ are defined as 
\begin{align}
   \mathcal{A}=\Box^2-\frac{3}{2}\frac{2+(1-\nu^2)}{\ell^2}\Box+\frac{9}{2}p^{\nu}\nabla_{\nu}(p^{\mu}\nabla_{\mu}), \\
    \mathcal{B}_{\alpha\beta}=\left(-\Box-\frac{2+(1-\nu^2)}{\ell^2}\right)g_{\alpha\beta}+\frac{3(1-\nu^2)}{\ell^2}p_{\alpha}p_{\beta}.
\end{align}
The ghost partition function is defined as the ratio of Jacobians, $\mathcal Z_{\rm ghost}=J_2/J_1$, which gives
\begin{equation}
    \mathcal Z_{\rm ghost}=\sqrt{\frac{\det(-\mathcal{A})_0\det(-\mathcal{B})_{1}}{\det(-\Box)_0}}.
\end{equation}
Hence, the TMG one-loop partition function around the type-$D_s$ background is given by
\begin{equation}
    Z_{\rm one-loop}^{\rm TMG}=\sqrt{\frac{\det(-\mathcal{A})_0\det(-\mathcal{B})_{1}}{\det(-\Box)_0 \;\det (\mathcal{L}+\mathcal{E})_{+} \;\det (\mathcal{L}-\mathcal{E})_{-}}}.
    \label{Eq:OneLoopPFTMG}
\end{equation}
Due to the presence of a non-trivial traceless Ricci tensor \eqref{eq:backgroundE}, the scalar determinants $\det(-\mathcal{A})_0$ and $\det(-\Box)_0$ do not simplify further, unlike in the simpler case of BTZ in TMG discussed in \cite{Gaberdiel:2010xv}. In addition, the vector determinant $\det(-\mathcal{B})_{1}$ associated with the ghost sector has a non-trivial coupling to the Killing vector $p^\mu$. Finally, the determinants $\det (\mathcal{L}\pm\mathcal{E})_{\mp}$ couple the trace mode $h$ and the transverse-traceless mode $\left(h^{\rm TT}\right)_{\mu\nu}$ in a non-trivial way. As a consequence, computing the full one-loop partition function is a daunting task that goes beyond the scope of this work. Our goal is more modest: we will focus only on the Schwarzian and rotational modes, for which the coupling $\frac{4h}{3}\left(h^{\rm TT}\right)_{\mu\nu}\bar{E}^{\mu\nu}$ vanishes.\\

\subsection{Perturbation Theory}\label{sec:PertTheory}
We aim to compute the leading correction in the black hole temperature $T$, as the system approaches extremality. This amounts to computing the first-order corrections to the eigenvalues and eigenmodes of the zero modes in the black hole throat by solving the following eigenvalue problem perturbatively:
\begin{equation}
    (\mathcal{L} h^{\rm TT})_{\mu\nu}=\lambda \left(h^{\rm TT}\right)_{\mu\nu},
\end{equation}
where the modes satisfy traceless-traverse conditions 
\begin{equation}
\label{eq:TTGauge}
    \left(h^{\rm TT}\right)^{\mu}{}_{\mu}=0,\quad \nabla^{\mu}\left(h^{\rm TT}\right)_{\mu\nu}=0.
\end{equation}
From now on, in order to simplify notation, we drop the label TT as we will only focus on these modes for the rest of the paper. Let us expand our different objects at first order in the perturbation theory
\begin{align}
    \lambda&=\bar{\lambda}+T \delta \lambda,\\
    h_{\mu\nu}&=\bar{h}_{\mu\nu}+T \delta h_{\mu\nu},\\
    \mathcal{L}&=\bar{\mathcal{L}}+T\delta\mathcal{L}.
\end{align}
The spectral problem becomes
\begin{equation}  
\delta\mathcal{L}\,\bar{h}_{\mu\nu}+\mathcal{L} ^{(0)}\delta h_{\mu\nu}=\delta \lambda \;\bar{h}_{\mu\nu}+\bar{\lambda}\delta h_{\mu\nu}.
\end{equation}
We assume that the zero modes are normalizable with the following convention
\begin{equation}
    \left\langle \bar{h}^{(n)}|\bar{h}^{(m)}\right\rangle =\delta_{m+n,0}. 
\end{equation}
Thus, to isolate the first-order correction to the eigenvalue $\lambda^{(0)}$, we take the inner product for the left with the zero mode $\bar{h}_{\mu\nu}$ and obtain 
\begin{eqnarray}
\label{eq:PerturbInter}
    \delta \lambda=\left\langle \bar{h}^{(-n)}\Big|\delta\mathcal{L}\;\bar{h}^{(n)}\right\rangle +\left\langle \bar{h}^{(-n)}\Big|\bar{\mathcal{L}}\;\delta h^{(n)}\right\rangle -\bar{\lambda}\left\langle \bar{h}^{(-n)}\Big|\delta h^{(n)}\right\rangle .
\end{eqnarray}
Typically, one assumes an adiabatic principle, according to which the unperturbed Hilbert space $\mathcal{H}$ remains unchanged under perturbation theory. In practice, this means that the eigenfunction correction $\delta h_{\mu\nu}$ is normalizable, so that one can implement the expansion
\begin{equation}
\label{eq:Adiabatic}
    \delta h_{\mu\nu} =\sum_n c_n \bar{h}_{\mu\nu}^{(n)},
\end{equation}
where $\bar{h}_{\mu\nu}^{(n)}$ is a complete basis. In this way, the last two terms in Eq. \eqref{eq:PerturbInter} cancel out. Hence, we obtain the standard formula for the eigenvalue correction 
\begin{eqnarray}
    \delta \lambda=\left\langle \bar{h}^{(-n)}\Big|\delta\mathcal{L}\;\bar{h}^{(n)}\right\rangle .
\end{eqnarray}
However, if the eigenfunction correction $\delta h_{\mu\nu}$ is non-normalizable\footnote{We thank Alejandra Castro for private discussions that clarified the non-normalizability of the rotational modes. These discussions motivated the prescription we adopt to address this technical subtlety within the black hole throat.}, the correct expression for the zero modes ($\bar{\lambda}=0$) is
\begin{eqnarray}
\label{eq:PerturbativeForm2}
    \delta \lambda=\left\langle \bar{h}^{(-n)}\Big|\delta\mathcal{L}\;\bar{h}^{(n)}\right\rangle +\left\langle \bar{h}^{(-n)}\Big|\bar{\mathcal{L}}\;\delta h^{(n)}\right\rangle ,
\end{eqnarray}
which holds under the assumption that, despite the fact that $\delta h_{\mu\nu}$ is non-normalizable, the following inner product is finite 
\begin{equation}
\label{eq:EnlargeH}
    \left\langle \bar{h}^{(-n)}\,\Big|\,\bar{\mathcal{L}}\;\delta h^{(n)}\right\rangle <\infty.
\end{equation}
The non-normalizable rotational modes do not belong to the original Hilbert space $\mathcal{H}$ of square-integrable perturbations relevant for the black hole throat. Thus, in perturbation theory, one must enlarge the space of admissible fluctuations in order to include non-normalizable modes which satisfy the condition \eqref{eq:EnlargeH}. Our prescription for computing the first-order correction to the eigenvalue
\eqref{eq:PerturbativeForm2} must take into account the fact that, on the
enlarged space of fluctuations, the operator $\bar{\mathcal{L}}$ is no longer
symmetric.\footnote{A symmetric operator $\mathcal{S}$ is defined as $\langle  n\,|\,\mathcal{S}m\rangle=\langle  m\,|\,\mathcal{S}n\rangle$ for any two arbitrary states $n,m\in \mathcal{H}$.}, due to the presence of a non-vanishing flux 
\begin{equation}
    f=\left\langle h_1\, \Big|\,\bar{\mathcal{L}}\, h_2 \right\rangle-\left\langle h_2 \,\Big|\,\bar{\mathcal{L}}\, h_1 \right\rangle\neq 0,
\end{equation}
with $h_{1,2}$ two arbitrary fluctuations. In the present context, the unperturbed eigenfunctions are zero modes and are therefore degenerate. Consequently, computing their first correction would normally require going to second order in perturbation theory. However, we can bypass this technical complication by exploiting the traceless-transverse conditions Eqs. \eqref{eq:TTGauge}, which must also hold order by order in the temperature expansion. For instance, the trace condition yields
\begin{equation}
\label{eq:TTgaugePerturbative}
    0= \left(h^{\rm TT}\right)^{\mu}{}_{\mu}=0+T\;\delta g^{\mu\nu}(\bar{h}^{\rm TT})_{\mu\nu}+T\;\bar{g}^{\mu\nu}(\delta h^{\rm TT})_{\mu\nu}+\mathcal O(T^2).
\end{equation}
Hence, terms that are linear in $T$ should compensate one another. A similar argument applies to the transverse condition. A technical difficulty is that the metric fluctuation contains more independent components than can be fixed by imposing the traceless-transverse conditions alone. To address this issue, we organize the metric fluctuations using the Newman--Penrose (NP) formalism. In terms of NP basis, we will see that the Schwarzian and rotational zero modes have a particular decomposition. Our working assumption is that, when an infinitesimal temperature $T$ is turned on, the NP decomposition of the Schwarzian and rotational zero modes is preserved. This effectively reduces the number of metric components that need to be considered. This allows us to compute explicitly the first correction to eigenfunctions $(\delta h^{\rm TT})_{\mu\nu}$ by imposing the traceless-transverse conditions. We present the details of these computations in the following sections. \\

\section{BTZ black holes in TMG}\label{sec:BTZinTMG}
In this section, we study the near-extremal physics of the BTZ black holes in Topologically Massive Gravity. The corresponding classical analysis was carried out in \cite{Castro:2019vog}, while its holographic-renormalization interpretation in the classical regime was developed in \cite{Castro:2025itb}. Here, we focus on the one-loop physics and, in particular, on the emergence of nearly-zero modes at very low temperature. We compute the corrections to the eigenvalues and eigenfunctions of the Schwarzian and rotational modes associated with the $sl(2,\mathbb{R})\times u(1)$ isometries of the near-horizon background.\\

A key subtlety arises for the rotational zero mode. As we show in Sec. \ref{sec:NHBTZ}, its eigenfunction correction is non-normalizable, and this contribution must be properly taken into account in order to obtain the correct eigenvalue correction. To validate this perturbative treatment, we compute in Sec. \ref{sec:FABTZ} the exact eigenvalues and eigenfunctions of the Schwarzian and rotational modes in the full BTZ geometry. Expanding the exact results to first order in temperature, we find perfect agreement with the throat computation. This confirms that the prescription described in Sec. \ref{sec:PertTheory} correctly captures the eigenvalue correction of the rotational mode.\\

\subsection{Classical Background} 
\label{Sec:ClassicalBgBTZ}
In Euclidean signature, one can write the line element of the BTZ black hole \cite{Banados:1992gq} as
\begin{align}
\label{eq:BTZmetric}
    ds^2=N(r)d\tau^2+N(r)^{-1}dr^2+r^2\left(d\phi+iN^{\phi}(r)d\tau\right)^2,
\end{align}
where the lapse and the shift are respectively
\begin{equation}
    N(r)=\frac{(r^2-r_{+}^2)(r^2-r_{-}^2)}{\ell^2r^2}, \qquad N^{\phi}(r)=\frac{r_{+}r_{-}}{\ell r^2},
\end{equation}
with $r_{+}$ and $r_{-}$ the event and inner horizon, respectively. In TMG, the surface charges are shifted with respect to AdS$_3$ gravity by the gravitational Chern--Simons term, such that the black hole mass and angular momentum are given by \cite{abbott1982stability,Deser:2003vh,Skenderis:2009nt}
\begin{align}
\label{eq:MJBTZ}
    M_{\rm TMG}&= \frac{ r_+^2+ r_-^2}{8 G\ell^2} - \frac{ r_+ r_-}{4G\mu \ell^3},\\
    J_{\rm TMG}&=   \frac{ r_+ r_-}{4G\ell} - \frac{ r_+^2+ r_-^2}{8G \mu \ell^2}~\label{Eq:JTMG}.
\end{align}
Our conventions are fixed by computing $T_{\rm TMG}$ via the standard Hawking relation $T_{\rm TMG}=\kappa/2\pi$ with $\kappa$ being the surface gravity. Besides, $\Omega_{\rm TMG}$ is computed via the horizon Killing vector $\vec{\xi}=\partial_t+\Omega_{\rm TMG}\partial_{\phi}$ such that $\xi^2=0$ at the event horizon. In addition, $M_{\rm TMG}$  and $J_{\rm TMG}$ are computed with respect to the Killing vectors $\partial_t$ and $\partial_{\phi}$, respectively. 
The absence of naked singularity and the assumption of positive energies imply that $\mu\ell\geq1$ \cite{Ciambelli:2020shy}. The conjugate variables for $M_{\rm TMG}$, $J_{\rm TMG}$ are respectively the temperature and angular velocity
\begin{equation}
    T_{\rm TMG}=\frac{r^2_{+}-r^2_{-}}{2\pi \ell^2r_{+}},\qquad \Omega_{\rm TMG}=\frac{r_{-}}{\ell r_{+}}.
\end{equation}
We notice that $T_{\rm TMG}$ and $\Omega_{\rm TMG}$ do not change with respect to AdS$_3$ gravity as these quantities are determined by regularity conditions on the Euclidean BTZ geometry. The corresponding Wald entropy is \cite{Solodukhin:2005ah, Tachikawa:2006sz, Park:1998qk} 
\begin{equation}
\label{eq:STMG}
    S_{\rm TMG}=\frac{\pi r_{+}}{2G}-\frac{\pi r_{-}}{2G\mu\ell }.
\end{equation}
Notice that so far, we have described the \textit{bulk} thermodynamics without referring to any dual field-theory interpretation, for which specific boundary conditions are required. We will come back to this point in due time.

\paragraph{Newman--Penrose basis.}\label{Sec:NPBasis} In this work, we will heavily use the Newman--Penrose formalism to organize fluctuations. The line element can be decomposed in the following way 
\begin{equation}
    ds^2=k\otimes l+l\otimes k+p\otimes p,
\end{equation}
where in Euclidean signature, the NP basis satisfies the usual conditions
\begin{equation}
    k^2=l^2=0, \qquad k\cdot l=1, \qquad p^2=1,\qquad p\cdot k=p\cdot l=0.
\end{equation}
In Lorentzian signature, $p$ is spacelike, while $k$ and $l$ are null. In Euclidean signature, $k$ and $l$ become complex one-forms with vanishing norm. There is, however, a freedom in the choice of Newman--Penrose basis. In our convention, we single out the Killing vector
\begin{equation}
    \vec{p}=\frac{1}{r_++r_-}\left(i\ell\partial_{\tau}-\partial_{\phi}\right),
\end{equation}
because, in the near-horizon limit, this direction becomes the $S^1$ fiber over AdS$_2$, see Eq. \eqref{eq:fiberBTZ}. Notice that this Killing vector is similar to, but distinct from, the horizon-generating Killing vector $\vec{\xi}$. In particular, $\vec{p}$ is unit-normalized everywhere, $p^2=1$, and therefore does not become null at the event horizon. By contrast, the horizon Killing vector satisfies $\xi^2\Big|_{r_{+}}=0$. \\

Once this choice is made, the two orthogonal null one-forms are fixed up to an overall normalization, which we choose symmetrically for both $k$ and $l$. For the BTZ metric in Eq. \eqref{eq:BTZmetric}, this leads to the following Newman--Penrose basis:
\begin{align}
\label{eq:BTZNP}
    k&=\frac{\sqrt{(r^2-r_{+}^2)(r^2-r_{-}^2)}}{\sqrt{2}(r_{+}+r_{-})}\left[\frac{i}{\ell}d\tau+ \frac{\ell r(r_{+}+r_{-})}{(r^2-r_{+}^2)(r^2-r_{-}^2)} dr-d\phi\right],\\
    l&=\frac{\sqrt{(r^2-r_{+}^2)(r^2-r_{-}^2)}}{\sqrt{2}(r_{+}+r_{-})}\left[-\frac{i}{\ell}d\tau+ \frac{\ell r(r_{+}+r_{-})}{(r^2-r_{+}^2)(r^2-r_{-}^2)} dr+d\phi\right],\\
    p&=\frac{1}{(r_++r_-)}
\left[
\frac{i}{\ell}
\left(
r^2-r_+^2-r_-^2-r_+r_-
\right)d\tau
-
\left(
r^2+r_+r_-
\right)d\phi
\right]\label{Eq:BTZNP2}.
\end{align}

\subsubsection{Boundary Conditions}\label{sec:BCbtz}

In this section we show two different sets of boundary conditions, the Brown--Henneaux ones \cite{Brown:1986nw} and the Compère--Song--Strominger (CSS) ones \cite{Compere:2013bya} and we discuss the consequences of the choice of boundary conditions at the level of the thermodynamics.\\ 

We recall that, in the AdS/CFT correspondence, the boundary metric is arbitrary but fixed. Otherwise, the dual CFT could not be placed on an arbitrary background, nor could one compute correlation functions of the boundary stress tensor by varying the metric source. This point is emphasized precisely in \cite{Skenderis:2009nt}. We note that asymptotically AdS spacetimes in TMG generally include irrational powers and, at $\mu\ell=1$, $\log r$-terms \cite{Henneaux:2009pw,Skenderis:2009nt,Skenderis:2009kd}, however, we restrict to integer powers, corresponding to the boundary graviton sector. For concreteness, let us choose the Fefferman-Graham (FG) gauge \cite{Fefferman:1985}
\begin{eqnarray}
    ds^2\Big|_{r}= \frac{r^2}{\ell^2}\left(g^{(0)}_{ab}+\frac{\ell^2}{r^2}g^{(2)}_{ab}\right)dx^adx^b+\dots
\end{eqnarray}
with $a,b\in\{\tau,\theta\}$ boundary indices, and the radial coordinate $r$ is held fixed. For Brown--Henneaux boundary conditions, one sets $g^{(0)}_{ab}=\eta_{ab}$ with $\eta_{ab}$ the flat metric. Then, the fluctuations obey the following asymptotic fall-offs in the bulk:\footnote{We can discuss Brown--Henneaux boundary conditions using different bulk gauge fixings. For instance, in the Bondi-like gauge employed in \cite{Castro:2025itb}, the allowed fluctuations behave as $h_{ab}\sim O(r)$. Despite this, the two copies of Virasoro algebras follow naturally. The key point is that following discussion should not depend on a particular gauge fixing.}
\begin{eqnarray}
\label{eq:BHfluctuations}
    g^{(2)}_{ab}\sim O(r^0). 
\end{eqnarray}
The analog of Brown--Henneaux boundary conditions for locally AdS$_3$ solutions in TMG were studied extensively in the literature \cite{Kraus:2005zm,Solodukhin:2005ns,Solodukhin:2006chern,Hotta:2008yq,Skenderis:2009nt}. The gravitational Chern--Simons term allows for unequal left- and right-moving central charges, $c_L\neq c_R$. In the dual CFT, this corresponds to a gravitational anomaly, equivalently described as a diffeomorphism or local Lorentz anomaly \cite{Kraus:2005zm,Solodukhin:2006chern}. The left and right central charges $c_{L/R}$ are given by 
\begin{equation}
    c_L=c_{\,\rm BH}\left(1+\frac{1}{\mu \ell}\right),\qquad  c_R=c_{\,\rm BH}\left(1-\frac{1}{\mu \ell}\right),
\end{equation}
parameterized in terms of the Brown--Henneaux central charge \cite{Brown:1986nw}
\begin{equation}
    c_{\,\rm BH}=\frac{3\ell}{2G}.
\end{equation}
This way, the entropy $S_{\rm TMG}$ can be understood as a CFT$_2$ entropy in terms of the central charges $c_{R/L}$ as follows \cite{Saida:1999ec,Solodukhin:2005ns,Kraus:2005vz,Sahoo:2006vz,Tachikawa:2006sz}   
\begin{equation}
    S_{\rm TMG}=\frac{\pi}{6\ell}\left(c_L(r_{+}-r_{-})+c_R(r_{+}+r_{-})\right).
\end{equation}
In the limit $\mu\to\infty$, the two central charges become equal and reduce to the Brown--Henneaux central charge. Consequently, the 
gravitational
entropy in TMG reduces to the standard BTZ entropy in AdS$_3$ gravity, 
\begin{equation}
    S_{\rm TMG}\to S_{\rm AdS_3}=\frac{\pi r_{+}}{3\ell}c_{\,\rm BH}.
\end{equation}
In the same limit, the standard BTZ thermodynamics of AdS$_3$ gravity is recovered, as follows from Eqs. \eqref{eq:MJBTZ} and \eqref{Eq:JTMG}. Therefore, our analysis of the eigenspectrum in the following sections also includes BTZ in AdS$_3$ gravity as the limiting case $\mu\to\infty$. Another commonly discussed limit in the literature is $\mu\ell\to1$, which corresponds to chiral gravity \cite{Li:2008dq,Gaberdiel:2010xv}. In this case, $\ell M_{\rm TMG}=|J_{\rm TMG}|$ despite the fact that $T\neq0$. In addition, at the chiral point, one has $c_L\neq0$ and $c_R=0$. In this work, we restrict to the range $\mu\ell>1$, although we briefly comment on the chiral point later when discussing the eigenvalue corrections in the black hole throat. \\

Another interesting set of boundary conditions relevant for this work are the CSS boundary conditions \cite{Compere:2013bya}. These are somewhat subtle, since they correspond to mixed boundary conditions: some components of the metric obey Dirichlet boundary conditions, while others obey Neumann boundary conditions. We begin by describing the boundary metric $g^{(0)}_{ab}$ through the following line element:
\begin{equation}
    d\tilde{s}^2=-dx^{+}dx^{-}+\partial_{+}\mathcal{P}\;(dx^{+})^2.
\end{equation}
Here, $x^{\pm}=t/\ell \,\pm\,\phi$  are the lightcone coordinates. We use a tilde to denote the boundary line element. In addition, $\partial_{+}\mathcal{P}(x^+)$ is an arbitrary but periodic function. The CSS boundary condition allows the following fluctuations 
\begin{eqnarray}
\label{eq:CSSfluctuations}
    g^{(0)}_{++}\sim O(r^2), \quad g^{(2)}_{++}\sim O(r^2),
\end{eqnarray}
while the following terms are held fixed
\begin{equation}
\label{eq:CSSfixed}
   g^{(0)}_{+-}\sim O(r^2), \quad g^{(2)}_{--}=\frac{4G}{\ell}\mathcal{P}_0\sim O(r^0).
\end{equation}
The analogous set of boundary conditions for BTZ in TMG were considered in \cite{Ciambelli:2020shy}. This implementation of the boundary conditions leads to the Virasoro--Kac--Moody algebra 
\begin{align}
\label{eq:KacmoodyBTZ}
    [\mathcal{L}_n,\mathcal{L}_m]&=(n-m)\mathcal{L}_{n+m}+\frac{c}{12}(n^3-n)\delta_{n+m},\\
     [\mathcal{P}_n,\mathcal{P}_m]&=\frac{\hat k}{2}n\delta_{n+m},\\
     [\mathcal{L}_n,\mathcal{P}_m]&=-m\mathcal{P}_{m+n},
\end{align}
with central charges \cite{Ciambelli:2020shy}
\begin{equation}
    c=\left(1+\frac{1}{\mu \ell}\right)c_{BH}, \quad \hat k=-4\mathcal{P}_0 \left(1-\frac{1}{\mu \ell}\right).
\end{equation}
The level $\hat k$ of the algebra is state-dependent. $\mathcal{L}_0$ and $\mathcal{P}_0$ are related to black hole parameters as follows
\begin{equation}
    \ell M_{\rm TMG}=\mathcal{L}_0+\mathcal{P}_0,\quad J_{\rm TMG}=\mathcal{P}_0-\mathcal{L}_0.
\end{equation}
Given these results, it is follows that $S_{\rm TMG}$ can be understood as the entropy of warped CFT in the corresponding Cardy limit \cite{Detournay:2012pc}
\begin{equation}
    S_{\rm TMG}=4\pi \Big(\sqrt{-\mathcal{P}_0\mathcal{P}_0^{\rm vac.}}+\sqrt{-\mathcal{L}_0\mathcal{L}_0^{\rm vac.}}\Big),
\end{equation}
with 
\begin{equation}
    \mathcal{L}_0^{\rm vac.}=-\frac{c}{24}, \quad \mathcal{P}_0^{\rm vac.}=-\frac{\ell}{16G}\left(1-\frac{1}{\mu \ell}\right).
 \end{equation}
The values of $\mathcal{L}_0^{\rm vac.}$ and $\mathcal{P}_0^{\rm vac.}$ can be derived by choosing the vacuum state to have $r_{+}=i\ell$ and $r_{-}=0$, which corresponds to global AdS$_3$ spacetime, as can be seen by plugging these values into the line element \eqref{eq:BTZmetric}.\\

We observe that, by construction, $\mathcal{P}_0$ is held fixed under the CSS boundary conditions \eqref{eq:CSSfixed}. Therefore, if $J_{\rm TMG}$ is also held fixed, the thermodynamic variables become overconstrained. This is why the grand canonical ensemble appears more natural than the canonical ensemble for CSS boundary conditions. Nevertheless, to facilitate a direct comparison with the existing BTZ literature, we work in the canonical ensemble in the following section. We discuss a generalized grand-canonical ensemble for the BTZ geometry in Appendix~\ref{sec:Appendix}, and for the warped BTZ geometry in Sec.~\ref{sec:WBTZ}.\\

In summary, the black hole entropy \eqref{eq:STMG} can be understood in different ways depending on the choice of boundary conditions. The CFT$_2$ dual description corresponds to Brown--Henneaux-like boundary conditions, while the WCFT dual description corresponds to CSS-like boundary conditions. We emphasize that, in principle, the choice of boundary conditions is different from the choice of thermodynamics ensembles. The latter can be thought as Legendre transformation between charges and chemical potentials without modifying the fall-offs of the corresponding bulk metric. However, this point is subtle for the CSS boundary conditions as we discussed above.  We will return to the discussion of boundary conditions.

\subsubsection{Near-extremal Limit} 

In what follows, we will work in an ensemble of fixed angular momentum, therefore we will assume Brown--Henneaux boundary conditions, as done in \cite{Castro:2019vog}. We report the grand canonical analysis of what follows in the appendix \ref{sec:Appendix}, which is compatible with CSS boundary conditions.\\

The extremal limit in the BTZ geometry is set by taking $r_+ = r_-\equiv r_0$. The extremal mass, entropy and angular momentum are \cite{Castro:2019vog}
\begin{equation}
\label{eq:fixedJ}
    M_{\rm ext}=\frac{r_0^2 c_R}{6\ell^3}, \quad S_{\rm ext}=\frac{\pi r_0c_R}{3\ell}, \quad  J_{\rm ext}=\frac{r_0^2 c_R}{6\ell^2}.
\end{equation}
We define the near-extremal by setting 
\begin{equation}
    \label{Eq:SeparationHorizon}r_{\pm}=r_0\pm\lambda \delta r-\frac{1+\mu\ell }{1-\mu\ell }\frac{(\lambda \delta r)^2}{2r_0},
\end{equation}
with $\lambda$ the decoupling parameter. The quadratic term $(\lambda \delta r)^2$ is needed to keep the angular momentum \eqref{eq:fixedJ} fixed. The leading departure in $T$ from extremality is 
\begin{align}
    M=M_{\rm ext}+\frac{T^2}{M_{\rm gap}},\\
    S=S_{\rm ext}+\frac{2T}{M_{\rm gap}}.
\end{align}
The scale at which semiclassical thermodynamical description breaks down due to large quantum fluctuations is determined by the mass gap 
\begin{align}
    M_{\rm gap}=\frac{12}{\pi^2\ell c_L}\,.
\end{align}
At the chiral point $\mu\ell =1$, the mass gap $M_{\rm gap}$ remains finite even though $M_{\rm ext}=J_{\rm ext}=S_{\rm ext}=0$ since $c_R=0$. This phenomenon resembles the near-extremal behaviour of hairy black holes in new massive gravity, studied in \cite{Acito:2026mmf}. The near-horizon, near-extremal limit is implemented at the level of the metric by the following coordinate transformations:
\begin{equation}
\label{eq:decouplingBTZ}
    \tau \to \frac{\ell^2}{4\lambda \delta r}\tau,\qquad r\to r_0 +\lambda r, \qquad \phi\to \phi-i\left(\Omega_0-\frac{2\lambda\delta r }{r_0 \ell}\right)\frac{\ell^2\tau}{4\lambda\delta r}.
\end{equation}
The combination $\lambda \delta r$ corresponds to 
\begin{equation}
    T_{\rm TMG}=\frac{\lambda\delta r}{2\pi\ell_2^2}.
\end{equation}
The low-temperature limit is obtained by slightly separating the horizons using Eq. \eqref{Eq:SeparationHorizon} and implementing the diffeomorphism \eqref{eq:decouplingBTZ}, sending then $\lambda\to0$. After defining $\cosh \eta=r/\delta r$, we obtain the near-horizon line element
\begin{equation}
\label{eq:NElineBTZ}
ds^2=\ell_2^2\left(\sinh^2\eta\, d\tau^2+d\eta^2\right)+r_0^2\left(d\phi+\frac{i\ell_2}{r_0}(1-\cosh\eta)d\tau\right)^2.
\end{equation}
In this coordinate system, the horizon is located at $\eta=0$, and the AdS$_2$ boundary is recovered when $\eta\to\infty$. Regularity at $\eta=0$ requires the Euclidean time coordinate $\tau$ to be $2\pi$-periodic. The effective AdS$_2$ radius $\ell_2$ is determined by the UV parameters as
\begin{equation}
    \ell_2=\frac{\ell}{2}.
\end{equation}
The NP basis for the near-horizon metric is 
\begin{align}
\label{eq:NENPBTZ}
    \bar{k}&=\frac{\ell_2}{\sqrt{2}}\left(i\sinh\eta d\tau+d\eta\right),\\  \bar{l}&=\frac{\ell_2}{\sqrt{2}}\left(-i\sinh\eta d\tau+d\eta\right),\\
    \bar{p}&=i\ell_2\left(1-\cosh\eta\right)d\tau+r_0d\phi.
\end{align}
As discussed in Sec. \ref{Sec:NPBasis}, the dual vector of $\bar{p}$ is the Killing vector
\begin{equation}
\label{eq:fiberBTZ}
    \vec{p}=\frac{1}{r_0}\partial_{\phi},
\end{equation}
describes the fiber direction $U(1)$ over the AdS$_2$ base manifold.  Beyond the leading near-horizon metric $\bar g_{\mu\nu}$, the decoupling limit also determines a first subleading correction. This gives the small-temperature expansion
\begin{equation}
    g_{\mu\nu}=\bar{g}_{\mu\nu}+T\delta g_{\mu\nu}+\mathcal{O}(T^2),
\end{equation}
with the irrelevant deformation $\delta g_{\mu\nu}$ given by\footnote{This expansion is not valid at the chiral point $\mu \ell_2=1/2$, where $\ell M_{\rm TMG}=|J_{\rm TMG}|$. This implies that, at the chiral point, the angular momentum cannot be held fixed in the same way as in our analysis. By allowing $J_{\rm TMG}$ to vary at first order, one could obtain the corresponding $\delta g_{\mu\nu}$ for chiral gravity. We will not work out this case explicitly, as it would not add any new physical insight for the purposes of this work.} 
\begin{align}
    \delta g_{\mu\nu}dx^\mu dx^\nu&= \frac{2\ell^4_2\pi(2+(1-2\mu \ell_2)\cosh\eta(-2+\cosh\eta))}{r_0(1-2\mu \ell_2)}d\tau^2+4\ell^2_2\pi r_0\cosh\eta d\phi^2\nonumber\\
    &+\frac{2\ell^4_2\pi(-2+(1-2\mu \ell_2)\cosh^2\eta)\coth\eta}{r_0(1-2\mu \ell_2) \sinh\eta}d\eta^2\nonumber\\
    &-\frac{i\ell^3_2\pi (5-2\mu\ell_2-8(1-2\mu \ell_2)\cosh\eta+(1-2\mu \ell_2)\cosh2\eta)}{(1-2\mu \ell_2)}d\tau d\phi.
\end{align}
The factor $\mu\ell_2$ in $\delta g_{\mu\nu}$ arises because we have fixed the angular momentum in the decoupling limit \eqref{Eq:SeparationHorizon}. Similar expansions have been presented in \cite{Kapec:2024zdj,Acito:2025hka,Bac:2026eqj} for BTZ in AdS$_3$ gravity. In fact, in the limit $\mu\to\infty$, we recover the irrelevant deformation $\delta g_{\mu\nu}$ presented in \cite{Kapec:2024zdj,Acito:2025hka,Bac:2026eqj} up to a radial diffeomorphism. Similarly, we record the subleading term for NP basis  
\begin{equation}
    k=\bar{k}+T\;\delta k,\quad l=\bar{l}+T\;\delta l,\quad p=\bar{p}+T\;\delta p,
\end{equation}
 given by

\begin{align}
\label{eq:TcorrectionNPBTZ1}
\delta k
&=
\frac{i\pi\ell_2^3}{\sqrt{2}\,r_0(1-2\mu\ell_2)}
\left[
(1-2\mu\ell_2)(\cosh\eta-2)
+(1+2\mu\ell_2)\frac{\cosh\eta}{\sinh^2\eta}
\right]\sinh\eta\,d\tau
\nonumber\\
&\quad
-
\frac{\pi\ell_2^3}{\sqrt{2}\,r_0(1-2\mu\ell_2)}
\left[
2-(1-2\mu\ell_2)\cosh^2\eta
\right]
\frac{\cosh\eta}{\sinh^2\eta}\,d\eta
-
\sqrt{2}\pi\ell_2^2\sinh\eta\,d\phi ,
\\
\delta l
&=
-\frac{i\pi\ell_2^3}{\sqrt{2}\,r_0(1-2\mu\ell_2)}
\left[
(1-2\mu\ell_2)(\cosh\eta-2)
+(1+2\mu\ell_2)\frac{\cosh\eta}{\sinh^2\eta}
\right]\sinh\eta\,d\tau
\nonumber\\
&\quad
-
\frac{\pi\ell_2^3}{\sqrt{2}\,r_0(1-2\mu\ell_2)}
\left[
2-(1-2\mu\ell_2)\cosh^2\eta
\right]
\frac{\cosh\eta}{\sinh^2\eta}\,d\eta
+
\sqrt{2}\pi\ell_2^2\sinh\eta\,d\phi ,
\\
\delta p
&=
\frac{i\pi\ell_2^3}{r_0(1-2\mu\ell_2)}
\left[
1-6\mu\ell_2
-4(1-2\mu\ell_2)\sinh^4\frac{\eta}{2}
\right]d\tau
+
2\pi\ell_2^2\cosh\eta\,d\phi .\label{eq:TcorrectionNPBTZ12}
\end{align}
At order $T$, the vector $\vec p$ remains Killing. This fixes the residual ambiguity in the choice of Newman--Penrose basis for the irrelevant deformation $\delta g_{\mu\nu}$. In the next section, this basis will provide a natural way to organize the metric fluctuations.\\

\subsection{Lichnerowicz Operator}
\label{Sec:LichneOpBTZ}

In the BTZ background, the trace of the Ricci tensor $E_{\alpha \beta}$ vanishes, such that the one-loop partition function Eq. \eqref{Eq:OneLoopPFTMG} takes the simple form
\begin{equation}
\label{eq:one-loopBTZ}
   \mathcal Z_{\rm one-loop}
    =
  \sqrt{
    \frac{
    \det\!\left(-\Box+2/\ell^2\right)^{\rm T}_1
    }{
    \det\!\left(
    \mathcal D^{\rm M}
    \left(-\Box-2/\ell^2\right)
    \right)^{\rm TT}_2
    }}.
\end{equation}
This expression was originally reported in \cite{Gaberdiel:2010xv}. An explicit way of writing the Lichnerowicz operator acting  on traceless-transverse modes is through
\begin{equation}
\label{eq:threeOperator}
    (\mathcal L h^{\rm TT})_{\mu\nu}
    =
    -\frac{1}{64\pi\ell^2}
    \left(
    \mathcal D^{\rm M}\mathcal D^{\rm L}\mathcal D^{\rm R}
    h^{\rm TT}
    \right)_{\mu\nu},
\end{equation}
with
\begin{equation}
    (\mathcal D^{\rm M})_{\mu}{}^{\alpha}
    \equiv
    \delta_{\mu}^{\alpha}
    +\frac{1}{\mu}\bar D_{\mu}{}^{\alpha},
    \qquad
    (\mathcal D^{\rm L/R})_{\mu}{}^{\alpha}
    \equiv
    \delta_{\mu}^{\alpha}
    \pm \ell\,\bar D_{\mu}{}^{\alpha}.
\end{equation}
Moreover, $\mathcal{D}^{\rm M}$, $\mathcal{D}^{\rm L/R}$ commute with each other. As a consequence, finding the kernel of the third-order Lichnerowicz operator is equivalent to finding those of three first-order operators. This is why for physical solutions, the operator $\mathcal{D}^{\rm M}$ projects onto the massive graviton $h_{\rm M}$ while the operators $\mathcal{D}^{\rm L/R}$ annihilates the two polarizations of the massless boundary gravitons  $h_{\rm R/L}$ \cite{Li:2008dq}. These two classes of solutions are conceptually distinct. Boundary gravitons are generated by large diffeomorphisms, namely $h=\mathcal{L}_{\xi}g$, while massive gravitons are not: they correspond to genuine propagating degrees of freedom in the bulk geometry. This distinction is important for our purposes, since we are interested in nearly-zero modes that emerge in the very-low-temperature regime. At $T=0$, both the massive graviton and the zero modes belong to the kernel of the Lichnerowicz operator. However, massive gravitons are physical solutions, whereas the zero modes of interest arise from redundancies of the gauge-fixed Lichnerowicz operator that become manifest only at $T=0$ associated with the $sl(2,\mathbb{R})\times u(1)$ isometries of the near-horizon background.
\\

Within the framework of this paper, the goal is to solve the following eigenvalue problem 
\begin{equation}
\label{eq:LichTTBTZ}
    \frac{1}{64\pi}
    \left(
    \delta_{\mu}^{\alpha}
    +
    \frac{1}{\mu}\bar D_{\mu}{}^{\alpha}
    \right)
    \left(
    -\Box-\frac{1}{\ell^2}
    \right)
    h_{\alpha\nu}^{\rm TT}
    =
    \lambda\, h_{\mu\nu}^{\rm TT}.
\end{equation}
Since the Lichnerowicz operator can be decomposed into three commuting derivative operators, there is an underlying auxiliary first-order problem 
\begin{equation}
\label{eq:Firstevp}
    \bar{D}_{\mu}{}^{\alpha}h_{\alpha\nu}^{\rm TT}=\gamma\;h_{\mu\nu}^{\rm TT}. 
\end{equation}
In this way, the physical eigenvalue $\lambda$ can be determined by the auxiliary eigenvalue $\gamma$ as follows 
\begin{equation}
\label{eq:lambdaBTZgamma}
    \lambda
    =
    \frac{1}{64\pi}
    \left(\frac{1}{\ell^2}-\gamma^2\right)
    \left(1+\frac{\gamma}{\mu}\right).
\end{equation}
The one-loop partition function \eqref{eq:one-loopBTZ} was previously computed in \cite{Gaberdiel:2010xv} with Dirichlet boundary conditions, more precisely with fluctuations of order $h^{\rm TT}_{\mu\nu}\sim O(r^0)$. The corresponding spectrum remains finite in the extremal limit, that is $T\to 0$. Instead, we will focus on the sector corresponding to vanishing eigenvalues as $T\to 0$. As we will discuss later, this condition will select a peculiar class of fluctuations with fall-off which is not fully compatible with the standard Dirichlet boundary conditions.\\ 

\subsection{Near-horizon Analysis}\label{sec:NHBTZ}
Extremal BTZ black holes in Einstein gravity admit zero modes of the Lichnerowicz operator\footnote{Notice that, compared to other cases considered in the literature of $\text{log}\, T$ corrections, the Lichnerowitz operator $\mathcal{L}^{\rm TMG}$ is third order in derivatives. A recent study of these corrections in the context of higher derivative theories can be found in \cite{Alvarado:2026kio,Acito:2026mmf}.} in the near-horizon geometry. These have already been reported in the literature \cite{Kolanowski:2024zrq,Kapec:2024zdj,Acito:2025hka,Bac:2026eqj}.  In this work, we show that these eigenfunctions are also zero modes of the TMG Lichnerowicz operator for BTZ, and we compute the first corrections to their eigenspectrum. To determine the eigenfunction corrections, we will now take advantage of the NP decomposition introduced in the previous section, which will prove to be an efficient way to impose the transverse-traceless gauge condition. Particular care will be required when extracting the eigenvalue correction of the rotational mode, as the corresponding eigenfunction correction is non-normalizable.\\

\subsubsection{Schwarzian Modes}
At extremality, when $T=0$, the AdS$_2$ factor of the geometry supports an infinite family of zero modes \cite{Camporesi:1994ga}, denoted by $h^{\rm Schw.}_{\mu\nu\, (n)}$, corresponding to boundary time reparametrisations. These fluctuations will be referred to as \textit{Schwarzian modes}. On the near-horizon background \eqref{eq:NElineBTZ}, their explicit form is
\begin{equation}
\label{eq:SchwarzianCoordBTZ}
\bar h^{\rm Schw.}_{\mu\nu\, (n)}dx^\mu dx^\nu
=
\mathcal N_1 e^{in\tau}
\frac{\tanh^{|n|}(\eta/2)}{\sinh^2\eta}
\left(
-\sinh^2\eta\,d\tau^2
+
2i\frac{|n|}{n}\sinh\eta\,d\tau d\eta
+
d\eta^2
\right).
\end{equation}
These modes can be decomposed with respect to the near-horizon NP basis given in Eq. \eqref{eq:NENPBTZ} as follows
\begin{equation}
\label{eq:NPSchwarzianBTZ}
\bar{h}^{\rm Schw.}_{\mu\nu\,(n)}=\bar{h}^{(n)}_{kk}\;\bar{k}_{\mu}\bar{k}_{\nu}+\bar{h}^{(n)}_{ll}\;\bar{l}_{\mu}\bar{l}_{\nu},
\end{equation}
with
\begin{align}
    \bar{h}_{kk}^{(n)}=\mathcal N_1e^{in \tau}\ \frac{n+|n|}{n\ell_2^2}\frac{\tanh^{|n|}(\eta/2)}{\sinh^2\eta },\\
    \bar{h}_{ll}^{(n)}=\mathcal N_1e^{in \tau} \frac{n-|n|}{n\ell_2^2}\frac{\tanh^{|n|}(\eta/2)}{\sinh^2\eta },
\end{align}
along with the normalization factor
\begin{equation}
    \mathcal N_1=\frac{\ell_2}{2\pi}\sqrt{\frac{|n|(n^2-1)}{r_0}}.
\end{equation} 
As already pointed out in \cite{Acito:2025hka}, the Schwarzian modes can be written in Kerr--Schild form. This is natural from the perspective of JT gravity. After dimensional reduction to two dimensions, the Newman--Penrose structure is captured entirely by a pair of null one-forms. Our convention for the normalization of the modes is 
\begin{equation}
    \left\langle h^{(n)}|h^{(m)}\right\rangle =\delta_{m+n,0}. 
\end{equation}
As the normalization factor $\mathcal N_1$ suggests, the Schwarzian mode is not defined for $n=0,\pm1$ as these values corresponds to $sl(2,\mathbb{R})$ isometries of AdS$_2$ factor of the background \eqref{eq:NElineBTZ}.\\

\paragraph{$\boldsymbol T$-corrections.} The Schwarzian zero modes admit the simple Newman--Penrose decomposition given in Eq. \eqref{eq:NPSchwarzianBTZ}. We therefore assume that the leading correction in $T$ preserves this tensorial structure, so that only the scalar NP components and the basis one-forms are corrected. In other words, our ansatz is given by 
\begin{align}
   {h}^{\rm Schw.}_{\mu\nu\, (n)}=&\bar{h}^{\rm Schw.}_{\mu\nu\, (n)}+T\delta {h}^{\rm Schw.}_{\mu\nu\, (n)}+\mathcal O(T^2)\nonumber\\
   =&\left(\bar{h}_{kk}^{(n)}+T\delta h_{kk}^{(n)}\right)\left(\bar{k}+T \delta k\right)_{\mu}\left(\bar{k}+T \delta k\right)_{\nu}+(k \leftrightarrow l)+\mathcal O(T^2)\,,
\end{align}
 The explicit expressions for $\delta k_{\nu}$ and $\delta l_{\nu}$ are given in Eqs. \eqref{eq:TcorrectionNPBTZ1}$-$\eqref{eq:TcorrectionNPBTZ12}. By construction, the Schwarzian mode is traceless. Imposing the traverse condition $\nabla_{\mu} {h}^{\rm Schw.}_{\mu\nu\, (n)}=0$, one can determine the order-$T$ corrections $\delta  h_{kk}$ and $\delta h_{ll}$, resulting in
\begin{equation}
\label{eq:TcorrectionEigenfunctionBTZ}
     \delta h^{(n)}_{kk}=\frac{\pi\ell^2_2 \bar{h}^{(n)}_{kk}}{r_0}F(\eta),\quad
     \delta h^{(n)}_{ll}=\frac{\pi\ell^2_2 \bar{h}^{(n)}_{ll}}{r_0}F(\eta),
\end{equation} 
where $F(\eta)$ is defined as 
\begin{equation}
    F(\eta)
=c_1+
\frac{
(1+2\mu\ell_2) |n|
-8\mu\ell_2\cosh\eta
-2(1-2\mu\ell_2)\cosh^3\eta
}{
(1-2\mu\ell_2)\sinh^2\eta
}+2|n|\log\sinh\eta.
\end{equation}
The corrected eigenfunction also satisfies the auxiliary first-order eigenvalue problem \eqref{eq:Firstevp} at order $T$, for arbitrary values of the constant $c_1$. Since this constant does not affect the eigenvalue correction, we set it to zero in what follows. We also note that the correction $\delta h^{\rm Schw}_{\mu\nu\,(n)}$ remains normalizable. One can verify the annihilation of the inner product
\begin{eqnarray}
    \left\langle \bar{h}^{(-n)}_{\rm Schw.}\,\Big|\,\bar{\mathcal{L}}\;\delta h^{(n)}_{\rm Schw.}\right\rangle =0.
\end{eqnarray}
Therefore, the perturbation does not take us outside the original Hilbert space, and the usual adiabatic assumption remains valid in this sector. Physically, the black hole throat effectively acts as a closed system for the Schwarzian mode. We can thus use the standard formula for the eigenvalue correction 
\begin{equation}
\label{eq:EigenvalueCorrSchwBTZ}
    \delta \lambda_{\rm Schw.}^{(n)}=\left\langle \bar{h}^{(-n)}_{\rm Schw.}\,\Big|\,\delta \mathcal{L}\;\bar{h}^{(n)}_{\rm Schw.}\right\rangle =\frac{|n| T}{32r_0}\left(1+\frac{1}{2\mu \ell_2}\right),\quad \forall\, |n|>1.
\end{equation}
In the limit $\mu\to\infty$, that is when the gravitational Chern--Simons term decouples, we recover the result for the eigenvalue correction in AdS$_3$ gravity reported previously in the literature \cite{Kolanowski:2024zrq,Kapec:2024zdj,Acito:2025hka,Bac:2026eqj}. We remark that the eigenfunction correction given in Eq. \eqref{eq:TcorrectionEigenfunctionBTZ}, which is a novel result of this work, also holds in the limiting case of  AdS$_3$ gravity as they were obtained by solving the transverse condition in the BTZ black hole. \\

From the eigenvalue correction \eqref{eq:EigenvalueCorrSchwBTZ} we can compute the contribution to the partition function in the throat geometry
\begin{eqnarray}
    \delta  \log \, \mathcal Z_{\rm throat} = 2 \left( -\frac12 \right) \sum_{n \geq 2} \delta \lambda^{\rm Schw.}_{(n)}  = \log \left(\prod_{n \geq 2} \frac{32 r_0}{n T} \left(1+\frac{1}{2\mu \ell_2} \right)^{-1} \right) .
\end{eqnarray}
The infinite product is evaluated via zeta function regularization, $\prod_{n \geq 2} \frac{k}{n} = \frac{1}{\sqrt{2\pi} k^{3/2}}$, so that we finally obtain
\begin{eqnarray}
    \delta  \log \, \mathcal Z_{\rm throat} = \log \left( \left( \frac{T}{r_0}\right)^{3/2} \frac{1}{256\sqrt{\pi}}\left(1+\frac{1}{2\mu \ell_2} \right)^{3/2} \right) \sim \frac32 \, \log\, T
\end{eqnarray}
recovering the $\frac32 \log T$ behavior typical of the Schwarzian corrections \cite{Iliesiu:2022onk}.

\subsubsection{Rotational Modes}
\label{Sec:RotModeNHBTZ}
At extremality, when $T=0$, the $S^1$ fiber over the base manifold AdS$_2$ supports an infinite family of zero modes \cite{Kolanowski:2024zrq}, denoted by $h^{\rm Rot.}_{\mu\nu\, (n)}$. These modes, which we refer to here as \textit{rotational modes}, are given by
\begin{align}
    h^{\rm Rot.}_{\mu\nu\, (n)}dx^{\mu}dx^{\nu}=\mathcal{N}_2 e^{in \tau}\tanh\Big(\frac{\eta}{2}\Big)^{|n|}&\left[(-2+|n|+\cosh\eta)d\tau^2+\frac{-|n|+\cosh\eta}{\sinh^2\eta}d\eta^2\right.\\ &\,\left.\notag+
    2i\frac{|n|-n^2}{n\sinh\eta}d\tau d\eta+2i\frac{r_0}{\ell_2}d\phi d\tau+2\frac{r_0|n|}{\ell_2 n\sinh\eta} d\phi d\eta\right]
\end{align}
The Newman--Penrose decomposition of these modes is\footnote{Our convention for symmetrization of indices is: $B_{(\mu\nu)}=\frac{1}{2}(B_{\mu\nu}+B_{\nu\mu})$. } 
\begin{align}
\label{eq:NProtBTZ}
\bar{h}^{\rm Rot.}_{\alpha\beta\, (n)}=\bar{h}^{(n)  }_{kk}\bar{k}_{\alpha}\bar{k}_{\beta}+2\bar{h}^{(n)}_{pk}\bar{k}_{(\alpha}\,\bar{p}_{\,\beta)}+\left(\bar{k}\leftrightarrow \bar{l}\,\right),
\end{align}
with
\begin{align}
 \bar{h}_{pk}^{(n)}&= i\,\mathcal N_2 e^{in \tau}\frac{|n|+n}{2n}\frac{\tanh^{|n|}(\eta/2)}{\sinh\eta},\quad
    \bar{h}_{kk}^{(n)}=-\frac{\sqrt{2}(n-\cosh\eta)}{\sinh\eta}\bar{h}^{(n)  }_{pk},\\
    \bar{h}_{pl}^{(n)}&=  i\,\mathcal N_2 e^{in \tau}\frac{|n|-n}{2n}\frac{\tanh^{|n|}(\eta/2)}{\sinh\eta},\quad
    \bar{h}_{ll}^{(n)}=-\frac{\sqrt{2}(n+\cosh\eta)}{\sinh\eta}\bar{h}^{(n)}_{pl},
\end{align}
along with the normalization factor
\begin{equation}
\mathcal N_2
=
\frac{1}{\sqrt2\pi\ell_2}
\sqrt{\frac{|n|}{r_0}} .
\end{equation}
These modes are not defined for $n=0$ as this corresponds to the compact $ u(1)$ isometry of the background \eqref{eq:NElineBTZ}.\\ 

\paragraph{$\boldsymbol T$-corrections.} We now compute the first-order correction in $T$ to both the eigenfunction and the eigenvalue of the rotational mode. Assuming an adiabatic principle as the temperature is slightly turned on, and guided by the NP decomposition of the rotational mode Eq. \eqref{eq:NProtBTZ}, we take the following ansatz for the first-order correction to the eigenfunction:
\begin{align}
   {h}^{\rm Rot.}_{\mu\nu\, (n)}=&\,\bar{h}^{\rm Rot.}_{\mu\nu\, (n)}+T\delta {h}^{\rm Rot.}_{\mu\nu\, (n)}+\mathcal O(T^2)\nonumber\\
   =&\left(\bar{h}_{kk}^{(n)}+T\delta h_{kk}^{(n)}\right)\left(\bar{k}+T \delta k\right)_{\mu}\left(\bar{k}+T \delta k\right)_{\nu}\nonumber\\
   +&\,2\left(\bar{h}_{kp}^{(n)}+T\delta h_{kp}^{(n)}\right)\left(\bar{k}+T \delta k\right)_{(\mu}\left(\bar{p}+T\delta p\right)_{\nu)}\nonumber\\
   +&\,(k\leftrightarrow l\,)+\mathcal O(T^2).
\end{align}
By construction, this ansatz is traceless, then by imposing the traverse condition $\nabla^{\mu}{h}^{\rm Rot.}_{\mu\nu\, (n)}=0$, we determine the unknown variables to derive the order-$T$ correction, which are 
\begin{align}
\label{eq:eigenFTMGU1}
    \delta h_{kk}^{(n)}
    &=
    \frac{\pi\ell_2^2}{r_0}\,
    \bar h_{kk}^{(n)}\,G_1(\eta),
    &
    \delta h_{pk}^{(n)}
    &=
    \frac{\pi\ell_2^2}{r_0}\,
    \bar h_{pk}^{(n)}\,G_2(\eta),
    \nonumber\\
    \delta h_{ll}^{(n)}
    &=
    \frac{\pi\ell_2^2}{r_0}\,
    \bar h_{ll}^{(n)}\,G_1(\eta),
    &
    \delta h_{pl}^{(n)}
    &=
    \frac{\pi\ell_2^2}{r_0}\,
    \bar h_{pl}^{(n)}\,G_2(\eta).
\end{align}
where $G_{1}(\eta)$ and $G_2(\eta)$ are defined as 
\begin{align}
G_1(\eta)
&=
c_1
+
\frac{
7+34\mu\ell_2
+8|n|(1+2\mu\ell_2)(|n|-3\cosh\eta)
+8\cosh 2\eta
+(1-2\mu\ell_2)\cosh 4\eta
}{
8(1-2\mu\ell_2)(|n|-\cosh\eta)\sinh^2\eta
}
\nonumber\\
&\quad
+
2|n|\log\sinh\eta,
\\
G_2(\eta)
&=
c_1
+
\frac{1+2\mu\ell_2}{1-2\mu\ell_2}
\frac{|n|-\cosh\eta}{\sinh^2\eta}
-\cosh\eta
+
2|n|\log\sinh\eta.
\end{align}
The eigenfunction correction also satisfies the auxiliary first-order eigenvalue problem \eqref{eq:Firstevp} at order $T$ for the arbitrary constant $c_{1}$, which can be set to zero, as their presence does not affect the eigenvalue correction. Note that the eigenfunction ${h}^{\rm Rot.}_{\mu\nu\, (n)}$ is non-normalizable
\begin{eqnarray}
    \left\langle \bar{h}^{(-n)}_{\rm Rot.}\,\Big|\,\delta h^{(n)}_{\rm Rot.}\right\rangle \to \infty,
\end{eqnarray}
which entails that the inner product
\begin{eqnarray}
    \left\langle \bar{h}^{(-n)}_{\rm Rot.}\,\Big|\,\bar{\mathcal{L}}\;\delta h^{(n)}_{\rm Rot.}\right\rangle \neq0,
\end{eqnarray}
does not vanish. This shows that the adiabatic principle does not hold in perturbation theory for the rotational modes. Consequently, the Hilbert space of the throat alone is not sufficient to capture the physics of the rotational mode. Additional information from the far-away region is required. In this sense, the black hole throat effectively acts as an open system for these modes. Thus, as we already emphasized in Sec. \ref{sec:PertTheory}, the correct way to compute the eigenvalue correction is
\begin{equation}
\label{eq:BTZrotLambaNH}
    \delta \lambda_{\rm Rot.}^{(n)}=\left\langle \bar{h}^{(-n)}_{\rm Rot.}\,\Big|\,\delta \mathcal{L}\;\bar{h}^{(n)}_{\rm Rot.}+\bar{\mathcal{L}}\;\delta h^{(n)}_{\rm Rot.}\right\rangle =-\frac{|n| T}{32r_0}\left(1-\frac{1}{2\mu \ell_2}\right),\qquad \forall\,|n|>0.
\end{equation}
We also note that the critical point $2\mu\ell_2=1$, where the TMG theory behaves as a logarithmic CFT \cite{Gaberdiel:2010xv}, the eigenvalue of the rotational mode vanishes. However, we emphasize that this limit cannot be taken at the level of the eigenfunction corrections \eqref{eq:eigenFTMGU1}, as this expression would blow up in this limit. Notice that, for the BTZ solution to be devoid of naked singularities, the  condition $2 \mu \ell_2 >1$ needs to be satisfied \cite{Ciambelli:2020shy}. As a consequence of this, the value of the correction found in Eq. \eqref{eq:BTZrotLambaNH} is negative, which signals a pathology in the background used as a saddle point. For the interval $2 \mu \ell_2 >1$ the corresponding WCFT dual  is  non-unitary \cite{Detournay:2012pc,Apolo:2018eky}. \\

The BTZ case in AdS$_3$ gravity was reported to give a vanishing result in \cite{Kolanowski:2024zrq,Bac:2026eqj} for the contribution of the rotational mode. In particular, Ref.~\cite{Bac:2026eqj} emphasized that, for the rotational mode, perturbation theory in the throat breaks down due to a violation of the adiabatic principle. In the AdS$_3$ gravity limit, that is $\mu\to \infty$, our result is instead non-vanishing, since we take particular care in computing the eigenvalue correction. The rotational and Schwarzian corrections presented here were computed using Brown--Henneaux boundary conditions at fixed angular momentum. We have also repeated the analysis with CSS boundary conditions and found the same results as those stated above, see appendix \ref{sec:Appendix}. In the following section, we compute the exact eigenvalue of the rotational mode in the full finite-temperature geometry. By expanding this result at small $T$, we verify that our perturbative prescription correctly reproduces the eigenvalue correction.

\subsection{Far-away Analysis}
\label{sec:FABTZ}
We now coming back to the full geometry analysis for which the relevant data are the metric Eq. \eqref{eq:BTZmetric} and Newman--Penrose basis Eqs. \eqref{eq:BTZNP}$-$\eqref{Eq:BTZNP2}. We solve the following spectral problem
\begin{equation}
\label{eq:LichTTBTZ2}
    \frac{1}{64\pi}\left(\delta_{\mu}^{\alpha}+\frac{1}{\mu}\bar{D}_{\mu}{}^{\alpha}\right)\left(-\Box-\frac{2}{\ell^2}\right)h_{\mu\nu}^{\rm TT}= \lambda \;h_{\mu\nu}^{\rm TT},
\end{equation}
looking for eigenfunctions which have vanishing eigenvalues in the extremal limit $T\to0$. As discussed in Sec. \ref{Sec:LichneOpBTZ}, this operator can be expressed in terms of three mutually commuting first-order operators. Consequently, there is an underlying first-order eigenvalue problem:
\begin{equation}
\label{eq:first-orderEigenvalue}
    \bar{D}_{\mu}{}^{\alpha}h_{\alpha\nu}^{\rm TT}=\gamma\;h_{\mu\nu}^{\rm TT},
\end{equation}
with the differential operator $\bar{D}_{\mu}{}^{\alpha}$ defined in \eqref{eq:curlD}. It follows that the Lichnerowicz eigenvalue $\lambda$ can be written in terms of the auxiliary eigenvalue $\gamma$ as 
\begin{equation}
    \lambda=\frac{1}{64\pi}\left(\frac{1}{\ell^2}-\gamma^2\right)\left(1+\frac{\gamma}{\mu}\right).
\end{equation}
Our strategy for the following computations is the following. We first use the Newman--Penrose decomposition of the Schwarzian and rotational modes as an ansatz for the full geometry, motivated by the corresponding throat analysis. We then solve the transverse condition for the graviton profiles, obtaining a particular solution together with homogeneous contributions. Finally, we fix the remaining homogeneous terms by requiring the resulting profiles to solve the first-order eigenvalue problem \eqref{eq:first-orderEigenvalue}. This condition is equivalent to, but simpler than, imposing the full Lichnerowicz eigenvalue equation \eqref{eq:LichTTBTZ2}.\\

\paragraph{Schwarzian modes.} Inspired by the throat decomposition of the Schwarzian modes \eqref{eq:NPSchwarzianBTZ}, our ansatz for these modes in the full geometry is 
\begin{equation}
\label{eq:FullSchwarzianmodeAdS}
   h^{\rm Schw.}_{\mu\nu}=h_{kk}\;k_{\mu}k_{\nu}+h_{ll}\;l_{\mu}l_{\nu}.
\end{equation}
By solving the transverse condition and the first-order eigenvalue problem \eqref{eq:first-orderEigenvalue}, we derive the following NP components
\begin{align}
    h_{kk}&=\tilde{\mathcal N}_1\frac{n+|n|}{2n}e^{i E_n \tau} f(r),\quad
    h_{ll}=\tilde{\mathcal N}_1\frac{n-|n|}{2n}e^{i E_n \tau} f(r),
\end{align}
where
\begin{equation}
    f(r)=(r^2-r_{+}^2)^{\frac{|n|}{2}-1}(r^2-r_{-}^2)^{-\frac{|n|}{2}\frac{r_{-}}{r_{+}}-1}, 
\end{equation}
and $E_n= 2\pi n T_{\rm TMG}$ the Matsubara frequencies. The normalization factor $\tilde{\mathcal N_1}$ is given by
\begin{equation}
\label{eq:NormBTZSCH}
    \tilde{\mathcal N}_1=\left(\frac{(r_{+}^2-r_{-}^2)^{4-|n|\left(1-\frac{r_{-}}{r_{+}}\right)}\Gamma\left(2+|n|\frac{r_{-}}{r_{+}}\right)}{2\ell^2\pi^2r_{+}\Gamma\left(|n|-1\right)\Gamma\left(3-|n|\left(1-\frac{r_{-}}{r_{+}}\right)\right)}\right)^{1/2},\qquad 1<|n|<\frac{3r_{+}}{r_{+}-r_{-}}.
\end{equation}
The eigenvalue of the first order problem \eqref{eq:first-orderEigenvalue} is 
\begin{equation}
    \gamma_{\rm Schw.}=\frac{1}{\ell}-\frac{|n|(r_{+}-r_{-})}{\ell r_{+}}. 
\end{equation}
We obtain the same Schwarzian modes $h^{\rm Schw.}_{\mu\nu}$ and eigenvalue $\gamma_{\rm Schw.}$ as those reported in \cite{Kolanowski:2024zrq,Acito:2025hka,Bac:2026eqj} for BTZ in AdS$_3$ gravity. In those works, the restricted range of $n$ required for convergence of the norm was already pointed out. In other words, the presence of the additional derivative operator $\left(\delta_{\beta}^{\alpha}+\frac{1}{\mu}\bar{D}_{\beta}{}^{\alpha}\right)$ does not affect the form of these modes. However, it modifies the exact eigenvalues relative to those of AdS$_3$ gravity as follows:
\begin{equation}
\label{eq:ExactEigenSchwBTZ}
    \lambda_{\rm Schw}^{(n)}=\frac{|n| (r_{+}-r_{-})\left(2r_{+}-|n|(r_{+}-r_{-})\right)}{64\ell^2 r_{+}^2 \pi} \left(1+\frac{r_{+}-|n|(r_{+}-r_{-})}{\mu \ell r_{+}}\right), \qquad \forall\,|n|>1.
\end{equation}
In the $\mu \to \infty$ limit, these eigenvalues reduce to those of AdS$_3$ gravity reported in \cite{Kolanowski:2024zrq,Acito:2025hka,Bac:2026eqj}. Finally, using the decoupling limit described in Eq. \eqref{Eq:SeparationHorizon}, we expand this spectrum at linear order in $T$
\begin{equation}
     \lambda_{\rm Schw}^{(n)}=\frac{|n| T}{32r_0}\left(1+\frac{1}{\mu \ell}\right)+\mathcal O(T^2),
\end{equation}
which precisely matches our perturbative result in the throat Eq. \eqref{eq:EigenvalueCorrSchwBTZ}.\\

\paragraph{Rotational modes.} Inspired by throat analysis of Sec. \ref{Sec:RotModeNHBTZ}, the ansatz for the rotational modes in the full geometry is 
\begin{equation}
\label{eq:FullRotmodeAdS}
   h_{\mu\nu}^{\rm Rot.}=h_{kk}\;k_{\mu}k_{\nu}+2h_{pk}\;p_{(\mu}k_{\nu)}+(k\leftrightarrow l).
\end{equation}
By solving the transverse condition and the first-order eigenvalue problem \eqref{eq:first-orderEigenvalue}, we derive the following NP components
\begin{align}
    h_{pk}&=\tilde{\mathcal N}_2e^{i E_n\tau}\frac{n+|n|}{2|n|} W_1(r), \quad 
    h_{kk}=\tilde{\mathcal N}_2e^{i E_n\tau}\frac{n+|n|}{2n} W_2(r),\\
    h_{pl}&=\tilde{\mathcal N}_2e^{i E_n\tau}\frac{n-|n|}{2|n|} W_1(r), \quad 
    h_{ll}=\tilde{\mathcal N}_2e^{i E_n\tau}\frac{n-|n|}{2n} W_2(r),
\end{align}
with the functions $W_{1,2}(r)$ given by
\begin{align}
    W_1(r)&=-\frac{i r_{+}}{\sqrt{2}(2r_{+}-|n|(r_{+}-r_{-}))}(r^2-r_{+}^2)^{\frac{|n|}{2}-\frac12} (r^2-r_{-}^2)^{-\frac{|n|}{2}\frac{r_{-}}{r_{+}}-\frac12}
    ,\\
    W_2(r)&=\frac{i(|n|(r_{-}-r_{+})(r^2+r_{+}r_{-})+r_{+}(2r^2-r_{+}^2-r_{-}^2))}{2\left(2r_{+}-|n|(r_{+}-r_{-})\right)}(r^2-r_{+}^2)^{-\frac{|n|}{2}-1}(r^2-r_{-}^2)^{-\frac{|n|}{2}\frac{r_{-}}{r_{+}}-1}.
\end{align}
The normalization factor $\tilde{\mathcal N_2}$ is 
\begin{equation}
\label{eq:NormBTZRot}
    \tilde{\mathcal N}_2
=
\left(
\frac{
2\left(2r_{+}-|n|(r_{+}-r_{-})\right)^2
\left(r_{+}^2-r_{-}^2\right)^{2-|n|\left(1-\frac{r_{-}}{r_{+}}\right)}
\Gamma\left(1+|n|\frac{r_{-}}{r_{+}}\right)
}{
\left(2r_{+}+|n|(r_{+}-r_{-})\right)
\ell^2\pi^2r_{+}^2
\Gamma\left(|n|\right)
\Gamma\left(1-|n|\left(1-\frac{r_{-}}{r_{+}}\right)\right)
}
\right)^{1/2}\,,
\end{equation}
with the following range for which the norm converges
\begin{equation}
     1<|n|<\frac{r_{+}}{r_{+}-r_{-}}.
\end{equation}
The eigenvalue of the first order problem \eqref{eq:first-orderEigenvalue} is 
\begin{equation}
    \gamma_{\rm Rot.}=-\frac{1}{\ell}-\frac{|n|(r_{+}-r_{-})}{\ell r_{+}}. 
\end{equation}
The existence of the rotational modes $h^{\rm Rot.}_{\mu\nu}$ was first observed in \cite{Acito:2025hka,Bac:2026eqj} for BTZ in AdS$_3$ gravity, where the restricted range of $n$ required for convergence of the norm was also reported. The presence of the additional derivative operator $\left(\delta_{\beta}^{\alpha}+\frac{1}{\mu}\bar{D}_{\beta}{}^{\alpha}\right)$ does not affect the form of these modes relative to AdS$_3$ gravity. However, it modifies the exact eigenvalues as follows:
\begin{equation}
\label{eq:ExactROtbtz}
    \lambda_{\rm Rot.}^{(n)}=\frac{|n| (r_{+}-r_{-})(|n|(r_{-}-r_{+})-2r_{+})}{64\ell^2 r_{+}^2\pi}\left(1+\frac{|n|(r_{-}-r_{+})-r_{+}}{ \mu\ell r_+}\right),\quad |n|>0.
\end{equation}
In the limit $\mu\to\infty$, this spectrum reduces to corresponding one in AdS$_3$ gravity \cite{Acito:2025hka,Bac:2026eqj}. At linear order in $T$, we obtain 
\begin{equation}
    \lambda_{\rm Rot.}^{(n)}=-\frac{|n| T}{32r_0}\left(1-\frac{1}{\mu \ell}\right)+\mathcal O(T^2),
\end{equation}
which is in exact agreement with Eq. \eqref{eq:BTZrotLambaNH}, obtained in the black hole throat region. \\

\subsection{Comments on Boundary Conditions} 
\label{sec:BCasymptotics}

In general, independently of the specific choice of boundary conditions, for asymptotically AdS spacetimes the boundary metric $g_{ab}^{(0)}$ appears at order $O(r^2)$ in the bulk metric expansion. With this in mind, we can now examine the asymptotic expansion of the Schwarzian modes \eqref{eq:FullSchwarzianmodeAdS} and rotational modes \eqref{eq:FullRotmodeAdS}. It is given by
\begin{align}
\label{eq:grow}
    h^{\rm Schw.}_{ab}&\sim O(r^m), \nonumber \\
     h^{\rm Rot.}_{ab}&\sim O(r^{2+m}), 
\end{align}
where we focused on the components $a,b=\{\tau,\theta\}$, as the remaining components are subleading. In order to simplify the discussion, we have introduced the exponent $m$ given by
\begin{equation}
    m=|n|\left(1-\frac{r_{-}}{r_{+}}\right). 
\end{equation} 
For the rotational modes, it is particularly useful to use light-cone coordinates, in which the expansion takes the form 
\begin{equation}
    h^{\rm Rot.}_{++}\sim O(r^{2+m}), \quad h^{\rm Rot.}_{+-}\sim O(r^{m}).
\end{equation} 
Notice that since $n$ is an integer, $m$ generally is a rational number. Boundary gravitons which are physical fluctuations have integer power of $r$. The most general possibility is to allow for an arbitrary boundary metric source. In this case, at finite temperature the exponent $m$ is bounded by $m\leq2$, which is equivalent to
\begin{equation}
\label{eq:m=2}
2\leq |n|\leq \frac{1}{\left(1-\frac{r_{-}}{r_{+}}\right)}.
\end{equation}
We emphasize that this choice \textit{differs} from the Brown--Henneaux boundary conditions \eqref{eq:BHfluctuations}, which are the standard choice of boundary conditions for computing the BTZ one-loop partition function in terms of the BTZ spectrum. \\

Now let us discuss the physical picture. Given that $|n| \geq 1$ for the Schwarzian modes, and $|n|>0$ for the rotational ones,  we see from Eq. \eqref{eq:grow} that at finite temperature $T$, the Schwarzian and rotational modes violate both the Brown--Henneaux boundary conditions \eqref{eq:BHfluctuations} and the CSS boundary conditions \eqref{eq:CSSfluctuations}, since they grow too rapidly near the AdS boundary. However, in the very low-temperature regime, where $r_{+}$ and $r_{-}$ coalesce, one has $m\to0$. In this limit, the Schwarzian fluctuations behave as $h^{\rm Schw.}_{ab}\sim O(r^0)$ consistently with Brown--Henneaux boundary conditions \eqref{eq:BHfluctuations}. In other words, the Schwarzian mode becomes normalizable in the throat, as already anticipated from the near-horizon analysis\footnote{For this discussion, it is important to recall that in the throat we introduce a new radial coordinate defined by $\cosh\eta=\frac{r}{\delta r}$.}. Notice that this picture arises when one takes the $r \rightarrow \infty$ limit first, and then the $T \rightarrow 0$ limit, not the other way around. \\

For the rotational mode, the situation is subtle. In the limit $m\to0$, these modes behave as $h^{\rm Rot.}_{ab}\sim O(r^{2})$, which is the expected asymptotic behaviour for sources. More precisely, in light-cone coordinates the asymptotic expansion is $h^{\rm Rot.}_{++}\sim O(r^{2})$ and $h^{\rm Rot.}_{+-}\sim O(r^{0})$, consistent with the CSS boundary conditions \eqref{eq:CSSfixed}. On the other hand, our near-horizon analysis showed that the rotational zero modes are normalizable. However, the first-order correction to the eigenfunctions, given in \eqref{eq:eigenFTMGU1}, is non-normalizable since it behaves as $\delta h^{\rm Rot.}_{\tau\tau}\sim O(r^{2})$, while all remaining components scale as $O(r^{0})$. This non-normalizable component is in a way ``inherited" from the analysis in the full geometry, albeit not in a straightforward way. In \eqref{eq:grow}, we first take the asymptotic limit $r\to\infty$ and only afterward the near-extremal limit $T\to0$. This ordering also obscures the distinction between the zero modes $\bar{h}^{\rm Rot.}_{ab}$, which are normalizable, and their first-order correction $\delta h^{\rm Rot.}_{ab}$, which is non-normalizable. Had the limits been taken in the opposite order, this distinction would have been manifest. This subtlety was discussed in detail from the perspective of holographic renormalization in \cite{Castro:2025itb}. There, it was shown that taking the limits in the order $T\to0$ followed by $r\to\infty$ fully captures the Schwarzian dynamics, whereas taking them in the opposite order, first $r\to\infty$ and then $T\to0$, leads to a subtle discrepancy at the level of the Weyl anomaly.\\

To recap, one can argue that the Schwarzian modes are compatible with both Brown--Henneaux and CSS boundary conditions, since as $T\to0$, these modes behave $h^{\rm Schw.}_{ab}\sim O(r^{0})$. By contrast, the rotational mode is compatible only with CSS boundary conditions as $T\to0$, since $h^{\rm Rot.}_{++}\sim O(r^{2})$ and $h^{\rm Rot.}_{-+}\sim O(r^{2})$, analogous to \eqref{eq:CSSfluctuations}.  This picture can be made more precise from the dual field theory perspective. In the CFT$_2$ description, which is compatible with Brown--Henneaux boundary conditions, \cite{Ghosh:2019rcj} showed that, in the near-extremal limit, the CFT$_2$ partition function leads to a logarithmic quantum correction to the entropy of $\frac{3}{2}\log T $ in both the canonical and grand canonical ensembles. Similarly, for the WCFT in the quadratic ensemble, which is compatible with CSS boundary conditions, \cite{Aggarwal:2022xfd} showed that, in the near-extremal limit, the WCFT partition function implies a logarithmic quantum correction to the entropy\footnote{The warped Schwarzian action was derived from the Bañados phase space of AdS$_3$ gravity with generalized CSS boundary conditions in \cite{Chaturvedi:2020jyy}. } of $\frac{3+1}{2}\log T $. The fact that the eigenvalue correction is negative precludes us to claim the matching with this WCFT result in the BTZ case, however we will find a more consistent picture when dealing with the WBTZ case. We will comment more on this in Section \ref{Sec:BCsWBTZ}.

\section{Warped Black Holes}\label{sec:WBTZ}
In this section, we study quantum effects in the near-extremal regime of the warped BTZ (WBTZ) black hole in topologically massive gravity \cite{Moussa:2003fc}. The classical analysis of this solution was carried out in \cite{Aggarwal:2023peg}, while its near-extremal description from the perspective of the warped CFT (WCFT) was developed in \cite{Aggarwal:2022xfd}. Here, we extend our analysis of the Schwarzian and rotational fluctuations, carried out in the previous section for the BTZ black hole in TMG, to the WBTZ black hole. We chose to work in the so-called \textit{quadratic ensemble}, whose extremal black hole admits a smooth BTZ limit \cite{Aggarwal:2023peg}. \\

Historically, warped $\mathrm{AdS}_3$ first appeared as the geometry of constant-$\theta$ slices of the near-horizon region of the Kerr black hole \cite{Bardeen:1999px}. This observation motivated the development of the Kerr/CFT correspondence \cite{Guica:2008mu} and, subsequently, the study of locally warped $\mathrm{AdS}_3$ spacetimes in three-dimensional theories of gravity such as TMG \cite{Anninos:2008fx}. In this sense, warped black holes provide an important toy model that captures several key physical features of Kerr black holes. One notable example is superradiance (see \cite{Brito:2015oca} for a review). Unlike the BTZ black hole, which does not exhibit superradiant modes \cite{Ortiz:2011wd}, the WBTZ black hole does support superradiant scattering \cite{Ferreira:2013zta}. Through this process, an extremal rotating black hole can be driven away from extremality. Consequently, superradiant modes cannot be neglected when studying quantum effects in near-extremal rotating black holes. Understanding their role remains one of the outstanding open problems in the physics of near-extremal black holes. A detailed study of superradiant modes in the low-temperature quantum regime lies beyond the scope of the present work and is left for future investigation.\\

In the present work we study the spectrum of the Lichnerowitz operator in WBTZ and we focus on constructing the exact finite-temperature profiles of the Schwarzian and rotational modes. We first show how to compute perturbative corrections to the Schwarzian and rotational-mode eigenvalues within the near-horizon throat, and then we extend the analysis in the full geometry. We elaborate on the subtleties regarding the (non-)normalizability of the rotational modes in the throat geometry, and demonstrate that they precisely reproduce the linear in temperature expansion of the exact result. \\

\subsection{Classical Background}
We consider the warped BTZ black hole \cite{Nutku:1993eb,Gurses:1994bjn, Bouchareb:2007yx, Anninos:2008fx}, a solution of TMG characterized by a non-vanishing Cotton tensor (or equivalently, a non-trivial traceless Ricci tensor). In Euclidean signature, the metric is given by
\begin{equation}
\label{eq:WBTZds}
    ds^2=N^2(r) dt^2+\frac{1-2 \mathrm{H}^2}{R^2(r)N^2(r)}r^2dr^2 + R^2(r) \Big(d\phi+iN^\phi (r)dt\Big)^2,
\end{equation}
with
\begin{align}
    R^2(r) &= (1-2\mathrm{H}^2)r^2 - 2 \mathrm{H}^2 \frac{(r^2-r_+^2)(r^2-r_-^2)}{(r_+ +r_-)^2},\\
    N^2(r)&= \frac{1-2\mathrm{H}^2}{R^2(r)L^2}(r^2-r_+^2)(r^2-r_-^2),\\
    N^\phi(r) &= \frac{1}{R^2(r)L}\left((1-2\mathrm{H}^2)r_+r_-+2\mathrm{H}^2 \frac{(r^2-r_+^2)(r^2-r_-^2)}{(r_++r_-)^2}\right),
\end{align}
where $r_{+}$ and $r_{-}$ are the event and inner horizon, respectively. One can relate $\mathrm{H}^2$ and $L$ to the TMG parameters $\mu$ and $\ell$ through
\begin{equation}
\label{eq:ParametersWbtz}
    L = \frac{2\ell}{\sqrt{\nu^2+3}},\quad
    \mathrm H^2 = -\frac{3}{2} \frac{\nu^2-1}{\nu^2+3}, \quad \nu=\frac{\mu\ell}{3}.
\end{equation}
The WBTZ black hole can also be described as a deformation of BTZ in the following manner\footnote{We mostly follow the conventions of \cite{Aggarwal:2023peg}, correcting a typo in the definition of the  unit-normalized Killing vector \eqref{eq:KillingPWbtz}. Moreover, our thermodynamic conventions differ from \cite{Aggarwal:2023peg} in two ways: it has an extra factor of $L$ in the denominator for $M, T,\Omega$, as well as an overall sign difference for $J$ and $\Omega$.}
\begin{equation}
    ds^2_{\rm WBTZ} = ds^2_{\rm BTZ}(L) -\frac{3+\nu^2}{2\nu^2} \mathrm{H}^2 (\nu) \,p\otimes p.
\end{equation}
The metric element $ds^2_{\rm BTZ}(L)$ stands for Eq. \eqref{eq:BTZmetric}, with an effective AdS radius $L$. In addition, the one-form $p$ is dual to the Killing vector
\begin{equation}
\label{eq:KillingPWbtz}
    \vec{p}=\frac{\sqrt{3+\nu^2}}{2\nu(r_{+}+r_-)}\Big(iL\partial_\tau-\partial_{\phi}\Big).
\end{equation}
As we already mentioned in Sec. \ref{Sec:LocallyAdSWBTZ}, this black hole solution is of type $D_s$ in the Petrov-Segre classification \cite{Chow:2009km}. In practice, this means that the Ricci traceless introduced in Eq. \eqref{eq:Edef} takes the form 
\begin{equation}
    E_{\mu\nu} = \frac{\nu^2-1}{\ell^2}\Big(g_{\mu\nu} - 3 p_\mu p_\nu \Big).
\end{equation}
In the limit $\nu \to 1$, that is when $\mathrm H \to 0$, the  traceless Ricci tensor $ E_{\mu\nu}$ vanishes. Therefore, the WBTZ solution reduces to the BTZ in TMG when the Chern--Simons coupling is fixed to $\mu \ell = 3$. Nonetheless, it is important to remind that BTZ in TMG is defined for arbitrary Chern--Simons coupling $\mu$. \\

\paragraph{Black hole Thermodynamics.} The WBTZ black hole is characterized by two charges: its mass and angular momentum
\begin{align}
    M=\frac{(3-4\mathrm H^2)( r_+^2+ r_-^2)-2 r_- r_+}{24G_3L^2\sqrt{1-2\mathrm H^2}},\qquad
     J=\frac{2(3-4\mathrm H^2) r_-  r_+ -( r_+^2+ r_-^2)}{24G_3L\sqrt{1-2\mathrm H^2}}.
\end{align}
The black hole temperature and angular velocity are
\begin{equation}
\label{eq:Twbtz}
    T=  \frac{r_+^2 - r_-^2}{2\pi r_+L^2},\qquad \Omega =  \frac{r_-}{L r_+},
\end{equation}
and the Wald entropy of this solution is
\begin{equation}
    S=\frac{\pi}{6G_3\sqrt{1-2\mathrm H^2}}\left((3-4\mathrm H^2)\,r_+-r_-\right).
\end{equation}
Given these quantities, we can verify the first law of thermodynamics
\begin{equation}
    dM=TdS+\Omega d J,
\end{equation}
as well as the \textit{Smarr formula}
\begin{eqnarray}
    2M=ST+2\Omega J.
\end{eqnarray}
Additionally, in the limit $\nu\to1$, the previous expressions match with the BTZ thermodynamics in TMG described in Sec. \ref{Sec:ClassicalBgBTZ} for the specific value of Chern--Simons coupling $\mu\ell=3$. So far, our discussion has focused on the \textit{bulk} thermodynamics, without invoking the dual field-theory description. The latter, however, depends on the choice of boundary conditions that we will discuss later.\\

\paragraph{Newman--Penrose Basis.} We can write down the Newman--Penrose triad that associated to the WBTZ metric \eqref{eq:WBTZds} as
\begin{align}
\label{eq:NPWBTZ}
    k&=\frac{\sqrt{(r^2-r_{+}^2)(r^2-r_{-}^2)}}{\sqrt{2}(r_{+}+r_{-})}\left(\frac{1}{L}id\tau+ \frac{rL(r_++r_-)}{(r^2-r_{+}^2)(r^2-r_{-}^2)} dr-d\phi\right),\\
    l&=\frac{\sqrt{(r^2-r_{+}^2)(r^2-r_{-}^2)}}{\sqrt{2}(r_{+}+r_{-})}\left(-\frac{1}{L}id\tau+ \frac{rL(r_++r_-)}{(r^2-r_{+}^2)(r^2-r_{-}^2)} dr+d\phi\right),\\
    p&=\frac{2\nu}{\sqrt{\nu^2+3}(r_++r_-)L}\Big(i\left(r^2-r_-^2-r_- r_+-r_+^2\right)d\tau-
L\left(r^2+r_- r_+\right)d\phi
\Big),
\end{align}
where $p$ is the one-form dual to the Killing vector $\vec p$ introduced in Eq.~\eqref{eq:KillingPWbtz}. The choice of this preferred Killing direction fixes a natural Newman--Penrose basis for the geometry.\\

\subsubsection{Warped CFT Algebra and Boundary Conditions} 
\label{sec:WCFT and phase space}
The WBTZ black hole is often referred to as the \textit{quadratic-ensemble} warped black hole, a terminology tied to the form of its asymptotic symmetry algebra: a Virasoro algebra semi-directly extended by a $\hat{ u}(1)$ Kac--Moody current \cite{Compere:2007in, Compere:2008cv, Compere:2009zj,Aggarwal:2020igb,Aggarwal:2023peg}:
\begin{align}
\label{eq:KacmoodyWBTZ}
    [\mathcal{L}_n,\mathcal{L}_m]&=(n-m)\mathcal{L}_{n+m}+\frac{c}{12}(n^3-n)\delta_{n+m},\\
     [\mathcal{P}_n,\mathcal{P}_m]&=\frac{\hat k}{2}n\delta_{n+m},\\
     [\mathcal{L}_n,\mathcal{P}_m]&=-m\mathcal{P}_{m+n},
\end{align}
with central charges
\begin{equation}
\label{eq:Centralk}
    c=\frac{2(1-\mathrm{H}^2)}{\sqrt{1-2\mathrm{H}^2}}\frac{L}{G}, \qquad \hat k=-4\mathcal{P}_0.
\end{equation}
The level $\hat k$ of the algebra is state-dependent, implying that the OPE $P(z)P(w)\sim \frac{\hat k(\mathcal{P}_0)}{(z-w)^2}$ is not completely determined by local data, since $\mathcal{P}_0$ is obtained from a closed integral of a local current. Consequently, the algebra is non-local. In the limit where $\nu\to1$, that is $\mu\ell\to3$, the WBTZ black hole reduces to the BTZ black hole with CSS boundary conditions discussed in Sec \ref{sec:BCbtz}. The central charges also match in this limit. Warped black holes can also be described in the so-called \textit{canonical ensemble}, which realizes the same asymptotic symmetry algebra \eqref{eq:KacmoodyWBTZ} but with a state-independent level $\hat k$. For further details, see \cite{Aggarwal:2022xfd,Aggarwal:2023peg}. In this work, we restrict to the quadratic ensemble. Throughout, we refer to the corresponding geometry as warped BTZ black holes.\\

The Virasoro--Kac--Moody algebra \eqref{eq:KacmoodyWBTZ} can be derived by the analysis of asymptotic symmetries in TMG \cite{Aggarwal:2020igb}. The corresponding boundary conditions in Fefferman--Graham gauge are given by 
\begin{equation}
    ds^2=\frac{L^2}{r^2}dr^2+g_{ab}(r,x)dx^adx^b,
\end{equation}
with the following fall-offs\footnote{This phase space does not include massive gravitons. To incorporate them, one would need to allow for non-integer powers of $r$ that depend on $\mu\ell $, as emphasized in \cite{Aggarwal:2020igb}.}
\begin{align}
\label{eq:falloffWBTZ}
    g_{++}&=r^4 j_{++}+r^2 h(x^+)+f_{++}(x^+)+O(r^{-1}),\\
    g_{+-}&=-\frac{\mu^2L^2}{18}r^2-\frac{h(x^+)}{j_{++}}\frac{(\mu^2L^2-9)}{36}+O(r^{-1}),\\
    g_{--}&=\frac{\mu^2L^2(\mu^2L^2-9)}{324 j_{++}}.
\end{align}
Here, we note that we are using $L$ given in \eqref{eq:ParametersWbtz} instead of the AdS radius $\ell$ of TMG. In addition, the light cone coordinates have been defined as $x^{\pm}=\tau/L\pm\phi$. This phase space can be viewed as the Bañados metrics for locally AdS$_3$ since it is locally WAdS$_3$. This phase space is characterized by three quantities: the constant $j_{++}$, the chiral functions $h(x^+)$, and $f_{++}(x^+)$. In the limit $\mu\ell\to3$, the phase space described above contains the asymptotically AdS spacetime in TMG with CSS boundary conditions. The WBTZ solution is included in this phase space and corresponds to the choice of functions
\begin{align}
    j_{++}&=\frac{\mu^2L^2-9}{72GL (LM-J)},\quad h=0,\\
    f_{++}&=\frac{GL}{9}(\mu^2L^2-9) (LM+J). 
\end{align}
The black hole mass and angular momentum can be related to the quantities in \eqref{eq:KacmoodyWBTZ} through
\begin{equation} \label{rel_alg_WBTZ}
    L M=\mathcal{L}_0+\mathcal{P}_0,\qquad J=\mathcal{P}_0-\mathcal{L}_0.
\end{equation}
We record here the value of $\mathcal{L}_0$ and $\mathcal{P}_0$ in terms of $r_+, r_-$:
\begin{equation} \label{rel_algL0}
    \mathcal{L}_0 = \frac{(1-H^2) (r_+-r_-)^2}{12G L \sqrt{1-2\mathrm{H}^2} } ,\qquad \mathcal{P}_0= \frac{\sqrt{1-2\mathrm{H}^2} (r_++r_-)^2}{24 G L} \,.
\end{equation}
For the purposes of this work, the most important feature of this phase space is the fall-offs given in \eqref{eq:falloffWBTZ}. For further details on how the charges $\mathcal{L}_n$ and $\mathcal{P}_n$ are obtained from the holomorphic functions $h(x^+)$ and $f_{++}(x^+)$, see \cite{Aggarwal:2020igb}.\\

Since the level $\hat k$ is determined in terms of $\mathcal{P}_0$ via \eqref{eq:Centralk}, the thermodynamic canonical ensemble in the near-extremal limit is somewhat ill-suited, as fixing both $\mathcal{P}_0$ and $J$ corresponds to keeping $\mathcal{L}_0$ fixed, thereby having no energy excitations, since the mass $M$ is fixed as well. For this reason, in what follows we will work in an ensemble where the angular momentum is allowed to fluctuate, while keeping $\mathcal{P}_0$ fixed.
 
\subsubsection{Near-extremal Limit}
We now proceed to implement the near-extremal limit in the grand canonical ensemble\footnote{In the CFT side,  $T_{L,R}$ are chemical potentials conjugate to $\mathcal L_0,\mathcal P_0$ and what is usually referred to as near-extremal \textit{grand canonical} ensemble corresponds to the limit $T_L \rightarrow 0$ and $T_R$ fixed. With a slight abuse of nomenclature, we also refer to our ensemble in gravity as \textit{grand canonical}.} in terms of the left- and right-moving temperatures, as done in \cite{Aggarwal:2023peg}. These temperatures are given by
\begin{equation}
    T_L=\frac{r_{+}-r_{-}}{2\pi L},\qquad T_R=\frac{r_{+}+r_{-}}{2\pi L}.
\end{equation}
Following the discussion in \cite{Aggarwal:2022xfd,Aggarwal:2023peg}, we fix the right-moving temperature $T_R$ and take an infinitesimal left-moving temperature $T_L \to 0$. The displacement of the horizon follows this formula:
\begin{equation} \label{displ_WBTZ}
    r_{\pm}=r_0\pm\lambda \delta r,
\end{equation}
with $\lambda$ the decoupling parameter. Notice that in this ensemble the angular momentum is allowed to fluctuate at order $(\lambda \delta r)^2$, hence we did not need a quadratic term in Eq. \eqref{displ_WBTZ}. The extremal values for the mass, entropy, and angular momentum are respectively:
\begin{equation}
    M_{\rm ext}=\frac{r_0^2\sqrt{1-2\mathrm{H}^2} }{6GL^2}, \quad S_{\rm ext}=\frac{\pi r_0\sqrt{1-2\mathrm{H}^2}}{3G}, \quad J_{\rm ext}=\frac{r_0^2\sqrt{1-2\mathrm{H}^2} }{6GL},
\end{equation}
while the departures in $T$ from extremality for the mass and entropy are respectively:
\begin{align}
    M=M_{\rm ext}+\frac{T^2}{M_{gap}},\\
    S=S_{\rm ext}+\frac{2T}{M_{gap}}. 
\end{align}
The breakdown scale at which quantum effects become more relevant than the semiclassical analysis is determined by the mass gap
\begin{align}
    M_{\rm gap}=\frac{6 G\sqrt{1-2\mathrm{H}^2}}{(1-\mathrm{H}^2)\pi^2L}=\frac{12 }{\pi^2 c}.
\end{align}
We now work out the near-horizon near-extremal limit in the geometry via the following coordinate transformations 
\begin{equation}
\label{eq:decouplingWBTZ}
    \tau \to \frac{L^2}{4 \lambda \delta r}\tau,\qquad r\to r_0 +\lambda \;r, \qquad \phi\to \phi+\frac{r_-}{r_+}\frac{L}{4\lambda\delta r }\tau,
\end{equation}
where the combination $\lambda \delta r$ is related to black hole temperature $T$ as follows 
\begin{equation}
    T=\frac{2\lambda\delta r}{\pi L^2}.
\end{equation}
Hence, the low-temperature limit is obtained by sending $\lambda\to0$. On top of that, by introducing the coordinate transformation $\cosh \eta=r/\delta r$, we obtain the near-horizon line element 
\begin{equation}
\label{eq:WBTZthroatMetricNice}
ds^2
=
\ell_2^2
\left(
d\eta^2+\sinh^2\eta\,d\tau^2
\right)
+
R_0^2
\left(
d\phi
+
\frac{i\ell_2}{r_0}(1-\cosh\eta)\,d\tau
\right)^2,
\end{equation}
where we introduced
\begin{equation}
\ell_2=\frac{\ell}{\sqrt{3+\nu^2}}=\frac L2,
\qquad
R_0=r_0\, C(\nu),
\end{equation}
and the warping is controlled by the parameter $\nu$ through 
\[
C(\nu) =\frac{2\nu}{\sqrt{3+\nu^2}}.
\]
In the unwarped limit $\nu\to1$, one has $C(\nu)\to1$, and the metric reduces to the standard self-dual AdS$_3$ geometry. For generic $\nu$, Eq.~\eqref{eq:WBTZthroatMetricNice} describes self-dual warped AdS$_3$, which is itself a solution of TMG \cite{Anninos:2008fx}. A convenient Newman--Penrose basis for the near-horizon metric is
\begin{align}
    \bar{k}
    &=
    \frac{\ell_2}{\sqrt{2}}
    \left(
    i\sinh\eta\, d\tau+d\eta
    \right),
    \\
    \bar{l}
    &=
    \frac{\ell_2}{\sqrt{2}}
    \left(
    -i\sinh\eta\, d\tau+d\eta
    \right),
    \\
    \bar{p}
    &=
    \frac{2\nu}{\sqrt{3+\nu^2}}
    \Big(
    i\ell_2(1-\cosh\eta)\,d\tau
    +r_0\,d\phi
    \Big).
\end{align}
In addition to the near-horizon metric, that is denoted by $\bar{g}_{\mu\nu}$, we will also need the subleading correction in the decoupling limit \eqref{eq:decouplingBTZ},
\begin{equation}
    g_{\mu\nu}=\bar{g}_{\mu\nu}+T\delta g_{\mu\nu}+\mathcal{O}(T^2),
\end{equation}
with  $\delta g_{\mu\nu}$ given by 
\begin{align}
    \delta g_{\mu\nu}dx^\mu dx^\nu =&
    \frac{2\pi \ell_2^4}{(3+\nu^2)r_0}
    (\cosh\eta-1)
    \Big(
    6(\nu^2-1)
    +(\nu^2+3)(\cosh\eta-1)
    -3(\nu^2-1)(\cosh\eta-1)^2
    \Big)d\tau^2 \nonumber\\
    &+
    \frac{4 i\pi \ell_2^3}{3+\nu^2}
    \Big(
    4\nu^2+2(3-\nu^2)(\cosh\eta-1)-(5\nu^2-3)(\cosh\eta-1)^2\Big)d\tau\,d\phi\nonumber\\
    &+\frac{2\pi \ell_2^4}{r_0}\cosh\eta\,d\eta^2
    +
    \frac{16\pi\nu^2\ell_2^2 r_0}{3+\nu^2}\cosh\eta\,d\phi^2.
\end{align}
As in the BTZ case, we expand the Newman--Penrose basis at first order in the temperature as
\begin{equation}
    \label{Eq:NPbasisWBTZ}
    k=\bar{k}+T\,\delta k,\qquad
    l=\bar{l}+T\,\delta l,\qquad
    p=\bar{p}+T\,\delta p .
\end{equation}
The corresponding first-order corrections are
\begin{align}
\delta k
&=
\frac{i\pi \ell_2^3}{\sqrt{2}\,r_0}
(\cosh\eta-2)\sinh\eta\, d\tau
+
\frac{\pi \ell_2^3}{\sqrt{2}\,r_0}
\cosh\eta\, d\eta
-
\sqrt{2}\pi \ell_2^2\sinh\eta\, d\phi,
\\
\delta l
&=
-\frac{i\pi \ell_2^3}{\sqrt{2}\,r_0}
(\cosh\eta-2)\sinh\eta\, d\tau
+
\frac{\pi \ell_2^3}{\sqrt{2}\,r_0}
\cosh\eta\, d\eta
+
\sqrt{2}\pi \ell_2^2\sinh\eta\, d\phi,
\\
\delta p
&=
\frac{2i\pi\nu \ell_2^3}{\sqrt{3+\nu^2}\,r_0}
\left(
2-(\cosh\eta-1)^2
\right)d\tau
+
\frac{4\pi\nu\ell_2^2}{\sqrt{3+\nu^2}}
\cosh\eta\, d\phi .
\end{align}
Here, $p$ is still taken such that its dual $\vec{p}$ is a Killing vector at order $T$. We will use these expansions to compute the first-order correction to eigenvalues and eigenfunction of zero modes of interest.\\

\subsection{Near-horizon Analysis}
\label{sec:ThroatAnalysis}
In this section, we compute the eigenvalue correction for the Schwarzian modes in extremal WBTZ. We further demonstrate that our prescription for the rotational mode reproduces the correct eigenvalue correction, upon comparison with the Taylor expansion of the exact result obtained in the following section.\\

\subsubsection{Schwarzian Modes}
At extremality, $T=0$, the near-horizon geometry contains Schwarzian zero modes. It is convenient to write these modes directly in the Newman--Penrose basis \eqref{Eq:NPbasisWBTZ}. We define the Schwarzian modes by
\begin{equation}
\label{eq:NPSchwarzianWBTZ}
\bar h^{\rm Schw}_{\mu\nu\,(n)}
=\bar h^{(n)}_{kk}\,\bar k_\mu \bar k_\nu+\bar h^{(n)}_{ll}\,\bar l_\mu \bar l_\nu ,
\end{equation}
with
\begin{align}
\bar h^{(n)}_{kk}
&=
\mathcal N_1 e^{in\tau}
\frac{n+|n|}{n\,\ell_2^2}
\frac{\tanh^{|n|}(\eta/2)}{\sinh^2\eta},\\
\bar h^{(n)}_{ll}
&=
\mathcal N_1 e^{in\tau}
\frac{n-|n|}{n\,\ell_2^2}
\frac{\tanh^{|n|}(\eta/2)}{\sinh^2\eta},
\end{align}
along with the normalization factor
\begin{equation}
\mathcal N_1
=
\sqrt{
\frac{
|n|(|n|^2-1)\ell_2^2
}{
4\pi^2 R_0
}
}.
\end{equation}

\paragraph{$\boldsymbol{T}$-corrections.} We now determine how the Schwarzian modes are lifted by the leading near-extremal deformation. As in the BTZ analysis, we assume that the Newman--Penrose structure of the mode is preserved at small temperature, while both the scalar NP components and the basis one-forms receive perturbative corrections. We therefore write
\begin{equation}
   h^{\rm Schw.}_{\mu\nu\,(n)}
   =
   \bar h^{\rm Schw.}_{\mu\nu\,(n)}
   +T\,\delta h^{\rm Schw.}_{\mu\nu\,(n)}
   +\mathcal O(T^2).
\end{equation}
In terms of the near-horizon Newman--Penrose basis, this expansion takes the form
\begin{align}
   h^{\rm Schw.}_{\mu\nu\,(n)}
   =
   &\left(\bar h_{kk}^{(n)}+T\,\delta h_{kk}^{(n)}\right)
   \left(\bar k+T\,\delta k\right)_\mu
   \left(\bar k+T\,\delta k\right)_\nu
   +(k \leftrightarrow l)
   +\mathcal O(T^2).
\end{align}
By construction, this expression is traceless. Solving the transverse condition at first order in $T$ fixes the correction to the scalar NP components up to an integration constant. The result is proportional to the extremal Schwarzian profile,
\begin{equation}
    \delta h_{kk}^{(n)}
    = \frac{2\pi \ell_2^2}{r_0}
    \bar h_{kk}^{(n)}\,F(\eta),
    \qquad
    \delta h_{ll}^{(n)}
    = \frac{2\pi \ell_2^2}{r_0}
    \bar h_{ll}^{(n)}\,F(\eta),
\end{equation}
with 
\begin{equation}
    F(\eta)
    =
    a_1
    -\cosh\eta
    +
    |n|\log\sinh\eta,
\end{equation}
with $a_1$ the integration constant. Plugging everything into the Lichnerowicz equation \eqref{eq:lichTMGTT}, we end up with the eigenvalue correction
\begin{equation}
\label{eq:EigenvalueCorrSchwWBTZ}
    \delta \lambda_{\rm Schw.}^{(n)}=\left\langle \bar{h}^{(-n)}_{\rm Schw.}\,\Big|\,\delta \mathcal{L}\;\bar{h}^{(n)}_{\rm Schw.}\right\rangle =\frac{  \left(5 \nu ^2+3\right) |n| T}{192  \nu ^2 r_0},\quad \forall\, |n|>1.
\end{equation}
In the limit $\nu\to1$, this result matches precisely with \eqref{eq:EigenvalueCorrSchwBTZ} evaluated at $2\mu\ell_2\to3$. \\

\subsubsection{Rotational Modes \label{rot_modes_WBTZ_NH}}

The near-horizon geometry also supports a second family of zero modes associated with the
$S^1$ fiber direction. We refer to these modes as rotational modes. In contrast with the
Schwarzian modes, their Newman--Penrose decomposition involves the spacelike one-form
$p$. At extremality, we write
\begin{align}
\label{eq:NPRotationalWBTZ}
\bar{h}^{\rm Rot.}_{\alpha\beta\, (n)}=\bar{h}^{(n)  }_{kk}\bar{k}_{\alpha}\bar{k}_{\beta}+2\bar{h}^{(n)}_{pk}\bar{k}_{(\alpha}\,\bar{p}_{\,\beta)}+(\bar{k}\leftrightarrow \bar{l}),
\end{align}
with
\begin{align}
\bar h^{(n)}_{kk}
&=
-\mathcal N_2 e^{in\tau}
\frac{(n+|n|)(n-\cosh\eta)}{n\,\ell_2^2}
\frac{\tanh^{|n|}(\eta/2)}{\sinh^2\eta},
\\
\bar h^{(n)}_{kp}
&=
\mathcal N_2 e^{in\tau}
\frac{\sqrt{3+\nu^2}}{2\sqrt{2}\,\nu\,n\,\ell_2^2}
(n+|n|)
\frac{\tanh^{|n|}(\eta/2)}{\sinh\eta},
\\
\bar h^{(n)}_{ll}
&=
\mathcal N_2 e^{in\tau}
\frac{(n-|n|)(n+\cosh\eta)}{n\,\ell_2^2}
\frac{\tanh^{|n|}(\eta/2)}{\sinh^2\eta},
\\
\bar h^{(n)}_{lp}
&=
-\mathcal N_2 e^{in\tau}
\frac{\sqrt{3+\nu^2}}{2\sqrt{2}\,\nu\,n\,\ell_2^2}
(n-|n|)
\frac{\tanh^{|n|}(\eta/2)}{\sinh\eta}.
\end{align}
The normalization factor is given by
\begin{equation}
\mathcal N_2
=
\sqrt{
\frac{
|n|\nu^2\,\ell_2^2
}{
2(\nu^2-3)\pi^2 R_0
}
}\,.
\end{equation}

\paragraph{$\boldsymbol{T}$-corrections.}
We now turn on the leading near-extremal deformation. As for the Schwarzian sector, we
keep the same Newman--Penrose tensor structure and allow both the scalar components
and the basis one-forms to receive first-order corrections. Thus
\begin{align}
   {h}^{\rm Rot.}_{\mu\nu\, (n)}=&\,\bar{h}^{\rm Rot.}_{\mu\nu\, (n)}+T\delta {h}^{\rm Rot.}_{\mu\nu\, (n)}+\mathcal O(T^2)\nonumber\\
   =&\left(\bar{h}_{kk}^{(n)}+T\delta h_{kk}^{(n)}\right)\left(\bar{k}+T \delta k\right)_{\mu}\left(\bar{k}+T \delta k\right)_{\nu}\nonumber\\
   +&\,2\left(\bar{h}_{kp}^{(n)}+T\delta h_{kp}^{(n)}\right)\left(\bar{k}+T \delta k\right)_{(\mu}\left(\bar{p}+T\delta p\right)_{\nu)}\nonumber\\
   +&\,(k\leftrightarrow l)+\mathcal O(T^2).
\end{align}
Solving the transverse condition at order $T$ fixes the NP component corrections as
\begin{align}
    \delta h_{kk}^{(n)}
    &=
    \frac{\pi \ell_2^2}{r_0}\,
    \bar h_{kk}^{(n)}\,G_1(\eta),
    \qquad
    \delta h_{pk}^{(n)}
    =
    \frac{\pi \ell_2^2}{r_0}\,
    \bar h_{pk}^{(n)}\,G_2(\eta),
    \nonumber\\
    \delta h_{ll}^{(n)}
    &=
    \frac{\pi \ell_2^2}{r_0}\,
    \bar h_{ll}^{(n)}\,G_1(\eta),
    \qquad
    \delta h_{pl}^{(n)}
    =
    \frac{\pi \ell_2^2}{r_0}\,
    \bar h_{pl}^{(n)}\,G_2(\eta),
\end{align}
with
\begin{align}
G_1(\eta)
&=a_2+
\frac{1+|n|^2}{|n|-\cosh\eta}-\cosh\eta
+
2|n|\log\sinh\eta+\frac{|n|}{2}\Big(1-\frac{3}{\nu^2}\Big),
\\
G_2(\eta)
&=a_2
-\cosh\eta
+
2|n|\log\sinh\eta .
\end{align}
In this computation, the homogeneous solution has been fixed by also imposing the auxiliary ``deformed'' first-order eigenvalue problem \eqref{eq:KeyEquation} at order $T$. Although this procedure does not determine the integration constant $a_2$, the final result is independent of its value. One can verify that norm of eigenfunction correction $\delta h^{\rm Rot.}_{\mu\nu}$ is non-normalizable. Thus, the correct way of computing the leading correction to rotational eigenvalue is
\begin{equation}
\label{eq:EigenvalueCorrRotWBTZ}
    \delta \lambda_{\rm Rot.}^{(n)}=\left\langle \bar{h}^{(-n)}_{\rm Rot.}\,\Big|\,\delta \mathcal{L}\;\bar{h}^{(n)}_{\rm Rot.}+\bar{\mathcal{L}}\;\delta h^{(n)}_{\rm Rot.}\right\rangle =\frac{  \left(\nu ^2-3\right) |n| T}{96  \nu ^2 r_0},\qquad \forall\,|n|>0.
\end{equation}
In the limit $\nu\to1$, this result matches precisely with \eqref{eq:BTZrotLambaNH} evaluated at $2\mu\ell_2\to3$. The correction has positive sign for $\nu>\sqrt3$. Notice that \cite{Aggarwal:2023peg,Anninos:2009zi} reported the BF bound $1 \leq \nu^2 \leq 15/11$ for the massive degree of freedom, which would make the eigenvalue correction negative, giving a divergent contribution to the path integral. \\

Notice that, when the eigenvalue correction is positive, we can compute the correction to the path integral due to the rotational mode:
\begin{eqnarray} 
    \delta \log \, \mathcal Z_{\rm throat} = 2 \left( -\frac12 \right) \sum_{n \geq 1} \delta \lambda^{\rm Rot.}_{(n)}  = \log \left(\prod_{n \geq 1} \frac{96 \nu^2 r_0}{n T (\nu^2-3)} \right),
\end{eqnarray}
which once again can be evaluated via zeta function regularization, obtaining
\begin{eqnarray} \label{corr_rot_quantum}
    \delta \log \, \mathcal Z_{\rm throat} = \log \left[ \frac{1}{8 \sqrt{3\pi}} \frac{\sqrt{\nu^2 -3}}{\nu}  \left( \frac{T}{r_0}\right)^{1/2}\right] \sim \frac12 \, \log\, T.
\end{eqnarray}
Hence, the rotational mode contributes with a factor $1/2\, \log\, T$ to $\mathcal Z_{\rm throat}$. \\

\subsection{Far-away Analysis}
We now coming back to the full geometry analysis where the relevant metric and its associated Newman--Penrose basis are given in Eqs. \eqref{eq:WBTZds} and  \eqref{eq:NPWBTZ} respectively. We want to solve the following spectral problem
\begin{equation}
\label{eq:lichTMGTT2}
    \left(\mathcal{L}h^{\rm TT}\right)_{\mu\nu}=\lambda(T) \;h_{\mu\nu}^{\rm TT},
\end{equation}
looking for eigenfunctions which have vanishing eigenvalues in the extremal limit $T\to0$. In this case, the Lichnerowicz operator $\mathcal{L}$ in the presence of non-trivial traceless Ricci tensor $\bar{E}_{\mu\nu}$
\begin{equation}
\label{eq:backgroundE2}
    \bar{E}_{\mu\nu} = \frac{\nu^2-1}{\ell^2}\left(g_{\mu\nu} - 3 p_\mu p_\nu \right),
\end{equation}
is rather involved and takes the following form 
\begin{align}
    \left(\mathcal{L}h^{\rm TT}\right)_{\mu\nu}&=\left(\delta_{\mu}^{\alpha}+\frac{1}{\mu}\bar{D}_{\mu}{}^{\alpha}\right)E^{(1)}_{\alpha\nu}[h^{\rm TT}]\nonumber\\
    &-\frac{\epsilon_{\mu}{}^{\alpha\beta}}{\mu}\Big(\bar{E}_{\beta\lambda}\Gamma^{\lambda(1)}_{\alpha\nu}[h^{\rm TT}]+\left(h^{\rm TT}\right)_{\alpha\lambda}\nabla^{\lambda}\bar{E}_{\beta\nu}+\left(h^{\rm TT}\right)_{\beta\lambda}\nabla_{\alpha}\bar{E}^{\lambda}{}_{\nu}\Big).
\end{align}
All the quantities entering  the Lichnerowicz operator are given in Sec. \ref{sec:WBTZLICh}. For asymptotically AdS$_3$ spacetimes, the TMG Lichnerowicz operator can be decomposed into three first-order operators, as in Eq. \eqref{eq:threeOperator}. Therefore, the first-order eigenvalue problem  introduced in Eq. \eqref{eq:first-orderEigenvalue} is guaranteed to exist. However, as already emphasized in \cite{Anninos:2009zi}, the presence of a non-trivial Cotton tensor spoils this simple structure. To handle this complication, we exploit two special features of the present setup:
\begin{enumerate}[$(i)$]
    \item the WBTZ black hole is a type-$D_s$ solution, so that Eq.~\eqref{eq:backgroundE2} holds,
    \item the Newman--Penrose formalism provides a basis adapted to this type-$D_s$ structure.
\end{enumerate}
Combining these two points with a suitable ansatz for the rotational and Schwarzian modes, we make an educated guess for a ``deformed" first-order eigenvalue problem for these special modes:
\begin{equation}
\label{eq:KeyEquation}
    D_{\mu}{}^{\alpha}h_{\alpha \nu}=\gamma\; h_{\mu\nu}+\beta(\nu)\; h_{kk}k_{\mu}k_{\nu},\\
\end{equation}
where the deformation parameter $\beta(\nu)$ is defined
such that $\beta(1)=0$ (which corresponds to the BTZ case). We stress that the last equation is not meant to hold for arbitrary perturbations. Rather, it applies only to the Schwarzian and rotational modes considered in this work. We attempted to formulate a more general, fully covariant first-order eigenvalue equation valid for arbitrary modes, but we were not able to identify such a structure. If it exists, it would likely provide a powerful route toward determining the full spectrum of the Lichnerowicz operator \eqref{eq:lichTMGTT2}. We leave this question as an open problem. We now proceed as in the BTZ analysis in TMG.\\

\paragraph{Schwarzian Modes.} Inspired by the throat analysis of Sec. \ref{sec:ThroatAnalysis}, our ansatz for the Schwarzian modes in the full geometry is 
\begin{equation}
\label{eq:FullSchwarzianmodeWBH}
   h^{\rm Schw.}_{\mu\nu}=h_{kk}\;k_{\mu}k_{\nu}+h_{ll}\;l_{\mu}l_{\nu}.
\end{equation}
By solving the transverse condition and the first-order eigenvalue problem \eqref{eq:KeyEquation}, we derive the following NP components for the Schwarzian eigenfunctions 
\begin{align}
h^{(n)}_{kk}
&=
\mathcal N_1
\frac{n+|n|}{2n}
e^{iE_n\tau}
f(r),
\\
h^{(n)}_{ll}
&=
\mathcal N_1
\frac{n-|n|}{2n}
e^{iE_n\tau}
f(r),
\end{align}
with
\begin{equation}
f(r)
=
\left(r^2-r_+^2\right)^{\frac{|n|}{2}-1}
\left(r^2-r_-^2\right)^{-1-\frac{|n|r_-}{2r_+}},
\qquad
E_n
=
2\pi n T .
\end{equation}
Here, $E_n$ is the Matsubara frequency related to the  WBTZ temperature given in Eq. \eqref{eq:Twbtz}. The normalization is
\begin{equation}
\label{eq:NormWBTSchRelation}
    \mathcal N_1
    =
    \alpha_1(\nu)\,
    \tilde{\mathcal N_1},
    \qquad
    \alpha_1(\nu)
    =
    \frac{(3+\nu^2)^{1/4}}{\sqrt{2\nu}} .
\end{equation}
The function $\alpha_1(\nu)$ is derived so that $\alpha_1(1)=1$, ensuring that the BTZ normalization is recovered in the unwarped limit. The factor $\tilde{\mathcal N_1}$ denotes the BTZ normalization in TMG, given in Eq. \eqref{eq:NormBTZSCH}. The eigenvalue and deformation parameter for the deformed first-order eigenvalue problem \eqref{eq:KeyEquation} are
\begin{equation}
    \gamma_{\rm Schw.}= \frac{\nu}{\ell}-\frac{|n|(r_{+}-r_{-})}{\ell r_{+}}\frac{3+\nu^2}{4\nu},\qquad \beta_{\rm Schw.}=0. 
\end{equation}
The eigenvalue $\gamma_{\rm Schw.}$ remains manifestly non-zero in the extremal limit $T\to0$. Moreover, apart from the overall normalization, the corresponding modes coincide with the Schwarzian modes found for BTZ in TMG. This can be understood from the fact that
\begin{equation}
    h^{\rm Schw.}_{\mu\nu} E^{\mu\nu}=0,
\end{equation}
so the Schwarzian perturbations do not couple to the traceless Ricci tensor that encodes the warping of the geometry. In this sense, the Schwarzian sector is orthogonal to the warped deformation, and its functional form is unchanged with respect to the BTZ case. Using these results, we can compute the eigenvalue correction 
\begin{equation}
    \lambda_{\rm Schw.}^{(n)}=\frac{ |n| T}{ r_{+}+r_{-}}\frac{3+4\nu\ell\gamma_{\rm Schw.}+\ell^2\gamma_{\rm Schw.}^2}{96\nu^2 }.
\end{equation}
This rather complicated expression in the limit $\nu\to1$ reduces to the corresponding eigenvalue Eq. \eqref{eq:ExactEigenSchwBTZ} for BTZ in TMG evaluated at $\mu\ell =3$. Furthermore, expanding this exact result at linear order in $T$, we obtain
\begin{equation}
    \lambda_{\rm Schw.}^{(n)}\simeq  \frac{  \left(5 \nu ^2+3\right) |n| T}{192  \nu ^2 r_0}+O(T^2).
\end{equation}
This result matches precisely with the perturbative result reported in Eq. \eqref{eq:EigenvalueCorrSchwBTZ} for WBTZ. In the limit, $\nu\to1$, this result also matches with the perturbative computation of BTZ in TMG reported in Eq. \eqref{eq:EigenvalueCorrSchwBTZ} evaluated at $\mu\ell=3$. \\

\paragraph{Rotational Modes.} 
We now turn to the full-geometry continuation of the rotational modes. Guided by the near-horizon analysis performed in Sec. \ref{sec:ThroatAnalysis}, we take the ansatz
\begin{equation}
\label{eq:FullRotmodeWBTZ}
   h_{\mu\nu}^{\rm Rot.}=h_{kk}\;k_{\mu}k_{\nu}+2h_{pk}\;p_{(\mu}k_{\nu)}+(k\leftrightarrow l).
\end{equation}
Solving the transverse condition together with the deformed first-order eigenvalue problem
\eqref{eq:KeyEquation}, we find
\begin{align}
    h_{pk}^{(n)}
    &=
    \mathcal N_2 e^{iE_n\tau}
    \frac{n+|n|}{2|n|}
    W_1(r),
    &
    h_{kk}^{(n)}
    &=
    \mathcal N_2 e^{iE_n\tau}
    \frac{n+|n|}{2n}
    W_2(r),
    \\
    h_{pl}^{(n)}
    &=
    \mathcal N_2 e^{iE_n\tau}
    \frac{n-|n|}{2|n|}
    W_1(r),
    &
    h_{ll}^{(n)}
    &=
    \mathcal N_2 e^{iE_n\tau}
    \frac{n-|n|}{2n}
    W_2(r).
\end{align}
The radial profiles are
\begin{align}
W_1(r)
&=\left(r^2-r_+^2\right)^{\frac{|n|-1}{2}}
\left(r^2-r_-^2\right)^{-\frac12-\frac{|n|r_-}{2r_+}}
,
\\
W_2(r)
&=
\frac{
\left(r^2-r_+^2\right)^{\frac{|n|}{2}-1}
}{
2\sqrt{2}\,\nu\sqrt{3+\nu^2}\,r_+}\left(r^2-r_-^2\right)^{-1-\frac{|n|r_-}{2r_+}}
\nonumber\\
&\times
\left[
8\nu^2 r_+\left(r_+^2-r^2\right)
+
(3+\nu^2)(r_- - r_+)|n|
\left(
-r^2+\frac{n}{|n|}r_+^2
\right)
\right].
\end{align}
The normalization is related to the BTZ one by a warping-dependent factor,
\begin{equation}
\label{eq:NormWBTZRot}
    \mathcal N_2
    =
    \alpha_2(\nu)\,
    \tilde{\mathcal N_2},
    \qquad
    1<|n|<\frac{r_+}{r_+-r_-}.
\end{equation}
The factor $\alpha_2(\nu)$ is 
\begin{equation}
     \alpha_2(\nu) =\sqrt{\frac{(3+\nu^2)^{3/2}(8r_{+}\nu^2+|n|(r_{-}-r_{+})(3+\nu^2))^2}{8(32r_{+}^2\nu^3(\nu^2-3)+|n|^2\nu(3+\nu^2)^2(r_{-}-r_{+})^2)}\frac{(-2r_{+}+|n|(r_{-}-r_{+}))}{(2r_{+}+|n|(r_{-}-r_{+}))}},
\end{equation}
such that $\alpha_2(1)=1$, so that the BTZ normalization is recovered in the unwarped limit.
The factor $\tilde{\mathcal N_2}$ denotes the BTZ normalization of the rotational modes given in Eq.~\eqref{eq:NormBTZRot}.
The deformed first-order eigenvalue problem is characterized by
\begin{align}
\label{eq:KeyEquationRotWBTZ}
\gamma_{\rm Rot.}
&=
-\frac{1}{\ell}
\left[
\nu
+
\frac{|n|(r_+-r_-)(3+\nu^2)}{4\nu r_+}
\right],
\\
\beta_{\rm Rot.}
&=
\frac{1}{\ell}
\frac{12\nu(\nu^2-1)r_+}
{
|n|(3+\nu^2)(r_- - r_+)+8\nu^2 r_+
}.
\end{align}
The deformation parameter vanishes in the unwarped limit $\nu\to1$, as expected from the BTZ result. For generic $\nu$, however, it is non-zero, reflecting the fact that the rotational perturbations couple to the warped part of the geometry:
\begin{equation}
    h_{\mu\nu}^{\rm Rot.}E^{\mu\nu}\neq0.
\end{equation}
Using these results, it is simpler to compute the eigenvalue of the Lichnerowicz operator \eqref{eq:lichTMGTT2} resulting in 
\begin{equation}
\label{eq:ExactRotEigenWBTZ}
\lambda_{\rm Rot.}^{(n)}= \frac{ |n|T}{ r_{+}+r_{-}}\frac{\gamma^2_{\rm Rot.} \ell^2+2 \nu\gamma_{\rm Rot.} \ell  +3 \nu ^2-6}{96\nu^2}.
\end{equation}
This rather complicated expression in the limit $\nu\to1$ reduces to the corresponding eigenvalue for BTZ in TMG \eqref{eq:ExactROtbtz} evaluated at $\mu\ell =3$. Furthermore, expanding this exact result at linear order in $T$, we obtain
\begin{equation}
    \lambda_{\rm Rot.}^{(n)}\simeq\frac{  \left(\nu ^2-3\right) |n| T}{96  \nu ^2 r_0}+O(T^2).
\end{equation}
This result matches precisely with the perturbative result reported in Eq. \eqref{eq:EigenvalueCorrRotWBTZ} for WBTZ. In the limit, $\nu\to1$, this result also matches with the perturbative computation of BTZ in TMG reported in Eq. \eqref{eq:BTZrotLambaNH}, when $\mu\ell=3$. 

\subsection{Comments on Boundary Conditions}
\label{Sec:BCsWBTZ}
The asymptotic expansion for the Schwarzian modes \eqref{eq:FullSchwarzianmodeWBH} and rotational modes \eqref{eq:FullRotmodeWBTZ} is given by
\begin{align}
\label{eq:grow2}
    h^{\rm Schw.}_{ab}&\sim O(r^m), \\
     h^{\rm Rot.}_{ab}&\sim O(r^{2+m}), \label{eq:grow3}
\end{align}
where we focus on the $a,b=\{\tau,\phi\}$ components as the remaining components are subleading.  For the rotational modes, it is particularly transparent to use light-cone coordinates $x^{\pm}=\tau/L\pm\phi$, in which the expansion takes the form 
\begin{equation}
    h^{\rm Rot.}_{++}\sim O(r^{2+m}), \quad h^{\rm Rot.}_{+-}\sim O(r^{m}).
\end{equation}
Notice that if we instead defined $x^{\pm}=\tau/\ell\pm\phi$, using the TMG AdS radius $\ell$ instead of $L$, the expansion would no longer take this simple form. In order to simplify the discussion we have introduced the exponent $m$ given by 
\begin{equation}
    m=|n|\left(1-\frac{r_{-}}{r_{+}}\right). 
\end{equation}
In Sec. \ref{sec:WCFT and phase space}, we discussed a phase space which is able to reproduce the Virasoro--Kac--Moody algebra \eqref{eq:KacmoodyWBTZ}, and it also contains the WBTZ black hole. The leading terms for the metric expansion are
\begin{equation}
\label{eq:boundryCondWBTZ}
    g_{++}\sim O(r^4), \quad g_{+-}\sim O(r^2), \quad g_{--}\sim O(r^0).
\end{equation}
In comparison to asymptotically locally AdS spacetimes, this phase space exhibits the unusual feature that some metric components fall off as $O(r^4)$. Nevertheless, for arbitrary integer values of $n$, the Schwarzian and rotational mode expansions in Eqs. \eqref{eq:grow2}$-$\eqref{eq:grow3} grow faster than the asymptotic behaviour allowed by this phase space as $r \to \infty$. Although TMG can admit irrational powers of the radial coordinate, such exponents are always bounded by an integer \cite{Aggarwal:2020igb,Henneaux:2009pw}. By contrast, the irrational exponent $m$ in Eq. \eqref{eq:grow2} becomes unbounded for fixed $r_{\pm}$ as $n$ increases. In summary, the Schwarzian and rotational modes at finite temperature are incompatible with the known physically reasonable set of boundary conditions in TMG.\\

In the very low-temperature regime $T \to 0$, corresponding to $m \to 0$, the discussion closely parallels that of Section \ref{sec:BCasymptotics} for BTZ in TMG. In this limit, the Schwarzian modes behave as $h^{\rm Schw.}_{ab} \sim O(r^0)$, while the rotational modes scale as $h^{\rm Rot.}_{++} \sim O(r^{2})$, and $h^{\rm Rot.}_{+-}\sim O(r^{0})$. Both therefore become compatible with the CSS boundary conditions \eqref{eq:boundryCondWBTZ} which leads to Virasoro--Kac--Moody algebra. As a consequence, the rotational mode must be included in the computation of the logarithmic correction to the WBTZ black hole entropy. Indeed for $\nu > \sqrt3$ the rotational eigenvalue correction is positive, leading to a $\frac12 \text{log}\, T$ correction as shown in \eqref{corr_rot_quantum}. The total contribution\footnote{With the caveat of the physical modes below the BF bound mentioned in Sec. \ref{rot_modes_WBTZ_NH}.} (Schwarzian plus rotational) is then $\frac{(3+1)}{2}\log T$. This result is consistent with the expected behaviour of the quadratic ensemble in the near-extremal limit of WCFT \cite{Aggarwal:2022xfd}.\\

Finally, the finite-temperature one-loop partition function around WBTZ has not yet been computed, as the relevant Lichnerowicz operator \eqref{eq:lichTMGTT2} is rather involved. As a result, whether the Schwarzian and rotational fluctuations should be included or excluded in such a one-loop analysis remains an open question. In this sense, the situation is less clear than in the BTZ case in TMG, where this computation can in principle be carried out explicitly.

\section{Conclusions and Discussion\label{conclusions}}

In this paper we have examined the spectral problem for the gravitational path integral in the background of BTZ and WBTZ black holes solution of Topologically Massive Gravity. We have discovered two infinite families of zero modes in the near-horizon geometry (Schwarzian and rotational zero modes), arising as diffeomorphisms left unfixed by the gauge choice in a higher-derivative gravity theory. We have regulated them by turning on temperature, and computed their eigenvalue correction. In addition, we have extended these modes to the full black hole geometry, giving rise to off-shell eigenmodes of the Lichnerowitz operator that satisfy Eq. \eqref{lichn}. \\

This serves our purpose of connecting the spectrum of zero-mode fluctuations in the horizon geometry to the asymptotic one, with a view of embedding the near-horizon picture into a boundary one. In this setup, we have found consistency between the one-loop contribution of these modes in the near-horizon geometry and that in the full spacetime. However, in order to achieve this one needs to carefully enforce the gauge condition in the near-extremal geometry, and in particular this entails including a non-normalizable correction to the eigenfunctions at the horizon. This non-normalizable eigenfunction correction for the rotational mode precisely matches a specific contribution of the modes in the full geometry, when they are subject to the decoupling limit in the  $T \rightarrow 0$ limit. Therefore, they can be interpreted as remnants of the gluing process to the asymptotic spacetime. Provided that one is careful in dealing with these subtleties, the agreement between near-horizon and far-away analysis holds for both BTZ and WBTZ.\\

A perhaps more puzzling point is the fact that, as anticipated in the introduction, the Schwarzian modes in the full BTZ geometry at finite $T$ have fall-offs that seem incompatible with standard Brown–Henneaux boundary conditions, as noticed in \cite{Acito:2025hka,Bac:2026eqj}, while in the very low-temperature regime they are. For warped BTZ black holes, an analogous subtlety arises: the Schwarzian and the rotational mode become compatible with CSS boundary conditions only in the very low temperature regime. Based on the near-horizon analysis, we find that the rotational mode contributes to the logarithmic correction to the WBTZ black hole entropy, yielding a total
contribution of $2\, \log \, T$ which matches the quadratic ensemble computations in the dual warped CFT \cite{Aggarwal:2022xfd}\footnote{In this regard, let us mention that Warped AdS$_3$/WCFT admits canonical and quadratic descriptions, related by a state-dependent change of variables. At the classical level, these descriptions can reproduce the same near-extremal thermodynamics, but their quantum low-temperature corrections differ. In the canonical
description, the fixed-angular-momentum partition function contains the warped-Schwarzian/JT factor and gives the expected $\frac{3}{2}\log T$ correction to the entropy. In the bulk, the extremal canonical ensemble solution is however not smoothly connected to the extremal BTZ black hole \cite{Aggarwal:2023peg}, and we have not considered it in the present work. In the quadratic description, by contrast, the corresponding low-temperature prefactor it predicts is $2 \log T$, in accordance with the inclusion of the rotational mode. }. All these observations make warped BTZ a sensitive probe of the near-extremal gravitational path integral. The one-loop answer can depend not only on the universal Schwarzian sector, but also on the fate of rotational modes, which boundary conditions are imposed, and which ensemble defines the path integral. Determining whether such modes are removed by matching to the full geometry is therefore essential for a consistent quantum thermodynamics of warped BTZ black holes.\\

Our analysis opens the way to different investigations. First of all, the Newman--Penrose formalism proved to be a formidable tool in finding analytic full geometry modes that extend the near-horizon ones. Our hope is that it will help facing the limitations that arise in higher-dimensional setups, for which full-geometry modes have been found numerically (see for example \cite{Kolanowski:2024zrq}), helping eventually to investigate how the Schwarzian is embedded in the boundary picture for CFT$_{\geq 2}$.  Along these lines, \cite{Castro:2021csm} studied linearized perturbation around near-NHEK using the NP formalism, making the extension of this analysis to the full Kerr geometry a natural next step. In a slightly different direction, it would also be interesting to reproduce the results we obtain here via the computation of the one-loop determinant via quasinormal modes \cite{Denef:2009kn,Denef:2009yy}: a careful investigation of the boundary conditions and the allowed ensemble are necessary for this purpose. In addition, the generalization of the work of \cite{Ferko:2024uxi} regarding AdS$_3$ string partition functions to the warped case would be interesting as well. We hope to report on these issues soon.\\

Lastly, we would like to point out that our zero-modes spectral analysis is not sufficient alone to determine the fate of the quantum corrections to black hole thermodynamics close to extremality. The WBTZ black hole supports superradiant modes (as well as the physical degrees of freedom mentioned in Sec. \ref{rot_modes_WBTZ_NH}), that are likely to render the quantum corrections studied here unobservable. A careful treatment of this point is left for future investigation but nonetheless is necessary for the full understanding the low temperature quantum regime behaviour.

\section*{Acknowledgements}

We thank  L. Acito, A. Castro, M. Lenzi  for useful discussions and collaboration on related topics. We are particularly grateful to the authors of \cite{Bac:2026eqj} for feedback on the issues related to the rotational mode. The work of RM and CT is supported
the MISU grant 40024018 ”Pushing horizons in Black hole Physics”. SD, RM and CT would like to thank the Galileo Galilei Institute for Theoretical Physics in Florence for hospitality during the final stages of this work. SD is a Senior Research Associate of the Fonds de la Recherche Scientifique F.R.S.-FNRS (Belgium).
SD acknowledges support of the Fonds de la Recherche Scientifique F.R.S.-FNRS (Belgium) through CDR n°40028632 (2025-2026). ED is a Research Fellow
of the Fonds de la Recherche Scientifique F.R.S.-FNRS (Belgium). The authors are member of BLU-ULB, the interfaculty research group focusing on space research at ULB. This work is supported by the F.R.S.-FNRS (Belgium) through convention IISN 4.4514.08 and benefited from the support of the Solvay Family.

\appendix

\section{Eigenfunction corrections for BTZ with CSS boundary conditions }
\label{sec:Appendix}
In the main text, we discussed the near-extremal limit of the BTZ black hole in TMG at fixed angular momentum $J_{\rm TMG}$. We present here the near-extremal analysis in the grand canonical ensemble, where the angular momentum is allowed to fluctuate. 

We start by introducing the left and right temperature 
\begin{equation}
    T_L=\frac{r_{+}-r_{-}}{2\pi \ell},\quad T_R=\frac{r_{+}+r_{-}}{2\pi \ell}.
\end{equation}
We fix the right-moving temperature $T_R$ and take an infinitesimal left-moving temperature $T_L \to 0$, which can be parameterized as follows:
\begin{equation}
    r_{\pm}=r_0\pm\lambda \delta r,
\end{equation}
The expansion for the BTZ metric \eqref{eq:BTZmetric} is performed by the following coordinate transformation 
\begin{equation}
    \tau \to \frac{\ell^2}{4 \lambda \delta r}\tau,\qquad r\to r_0 +\lambda \;r, \qquad \phi\to \phi+\frac{\ell}{4\lambda \delta r }\tau,
\end{equation}
The irrelevant deformation $\delta g_{\mu\nu}$ obtained is given by 
\begin{align}
    \delta g_{\mu\nu}dx^\mu dx^\nu =&
    \frac{2\pi \ell_2^4}{r_0}
    (\cosh\eta-1)^2
  d\tau^2 +
  2i\pi \ell_2^3
    \Big(
    2+2(\cosh\eta-1)-(\cosh\eta-1)^2\Big)d\tau\,d\phi\nonumber\\
    &+\frac{2\pi \ell_2^4}{r_0}\cosh\eta\,d\eta^2
    +4\pi\ell_2^2 r_0\cosh\eta\,d\phi^2.
\end{align}
From here the Newman--Penrose (NP) basis can be read off. The NP components for the first-order correction of the Schwarzian eigenfunctions are
\begin{align}
    \delta h_{kk}^{(n)}
    &= \frac{2\pi \ell_2^2}{r_0}
    \bar h_{kk}^{(n)}\,F(\eta),
    \qquad
    \delta h_{ll}^{(n)}
    = \frac{2\pi \ell_2^2}{r_0}
    \bar h_{ll}^{(n)}\,F(\eta),\\
    F(\eta)
    &=
    a_1-\cosh\eta+|n|\log\sinh\eta.
\end{align}
The NP components for the first-order correction of the Rotational eigenfunctions are
\begin{align}
    \delta h_{kk}^{(n)}
    &=
    \frac{\pi \ell_2^2}{r_0}\,
    \bar h_{kk}^{(n)}\,G_1(\eta),
    \qquad
    \delta h_{pk}^{(n)}
    =
    \frac{\pi \ell_2^2}{r_0}\,
    \bar h_{pk}^{(n)}\,G_2(\eta),
    \nonumber\\
    \delta h_{ll}^{(n)}
    &=
    \frac{\pi \ell_2^2}{r_0}\,
    \bar h_{ll}^{(n)}\,G_1(\eta),
    \qquad
    \delta h_{pl}^{(n)}
    =
    \frac{\pi \ell_2^2}{r_0}\,
    \bar h_{pl}^{(n)}\,G_2(\eta),
\end{align}
with the functions $G_{1,2}$ given by
\begin{align}
G_1(\eta)
&=a_2+
\frac{(1+|n|^2)}{|n|-\cosh\eta}-\cosh\eta
+
2|n|\log\sinh\eta-|n|,
\\
G_2(\eta)
&=a_2-\cosh\eta+2|n|\log\sinh\eta .
\end{align}
The integration constants $a_{1,2}$ do not play any role in computing the leading correction for the eigenvalues. Given these expressions for the eigenfunctions corrections, one obtains the exactly the same eigenvalues reported in Sec. \ref{sec:NHBTZ}. 

\bibliographystyle{JHEP}
\providecommand{\href}[2]{#2}\begingroup\raggedright\endgroup

\end{document}